\PassOptionsToPackage{protrusion, expansion}{microtype} %
\PassOptionsToPackage{dvipsnames}{xcolor} %
\pdfoutput=1
\documentclass[conference, nofonttune, 9pt]{IEEEtran}\def\UseIEEETemplate{1}  %
\IEEEoverridecommandlockouts%

\usepackage{graphicx}
\usepackage{xcolor}
\definecolor{B}    {HTML}{2b66d3}   %
\definecolor{B2}   {HTML}{003399}   %
\definecolor{Bv}   {HTML}{0000EB}   %
\definecolor{R}    {HTML}{c9171e}
\definecolor{R2}   {HTML}{d7003a}
\definecolor{INK}  {HTML}{595857}
\definecolor{Y}    {HTML}{f1c40f}
\definecolor{G}    {HTML}{009a00}
\definecolor{GRAY} {HTML}{808080}
\definecolor{MAUVE}{HTML}{9400D1}

\newcommand{\cusz}{{\scshape cuSZ}}
\newcommand{\cuszinterp}{\mbox{\textsc{cuSZ}-$i$}}
\newcommand{\thiswork}{\cuszinterp}
\newcommand{\ginterp}{\mbox{\textit{G-Interp}}}

\newcommand{\bitcomp}{Bitcomp-lossless}
\newcommand{\datasetname}[1]{{\ttfamily #1}}
\newcommand{\datafieldname}[1]{{\ttfamily #1}}

\usepackage{array}      %
\usepackage{booktabs}
\usepackage{colortbl}
\usepackage{adjustbox}
\usepackage{multirow}
\usepackage{tabu}       %

\usepackage[T1]{fontenc}

\usepackage{url, hyperref}                  %
\hypersetup{
    colorlinks, linkcolor=RoyalBlue, citecolor=RoyalBlue, urlcolor=RoyalBlue,
    allcolors=RoyalBlue,
    breaklinks=true,
    pdfborder={0 0 0}}

\usepackage{enumitem}                       %
\usepackage{lipsum}                         %

\usepackage{listings}						%
\usepackage{setspace}                       %

\usepackage{soul}%
\ifx\UseACMTemplate\undefined
\else
    \newcommand{\SEC}{\textcolor{black}{\S\hbox{\kern 0.1667em}}}
    \newcommand{\FIG}{\textcolor{black}{Figure~}}
    \newcommand{\TAB}{\textcolor{black}{Table~}}
\fi

\ifx\UseIEEETemplate\undefined
\else
    \newcommand{\SEC}{\textcolor{black}{\S\hbox{\kern 0.1667em}}}
    \newcommand{\FIG}{\textcolor{black}{Fig.\hbox{\kern 0.1667em}}}
    \newcommand{\TAB}{\textcolor{black}{TABLE~}}
\fi

\newcommand{\quantcode}{quant-code}

\colorlet{HLCOLOR}{B}
\colorlet{TableAltColor}{gray!20}

\newcommand{\FloatBodyStyle}{\centering\small\fontfamily{lmr}\selectfont}
\newcommand{\NoBcBestCr}[1]{\color{RoyalBlue} \usefont{OT1}{cmr}{bx}{n} {#1}}
\newcommand{\BestCr}[1]{\color{RoyalBlue} \usefont{OT1}{cmr}{bx}{n} {#1}}
\newcommand{\SecondbestCr}[1]{\color{black} \underline{#1}}

\usepackage{tikz}
\usetikzlibrary{positioning}        %
\usetikzlibrary{shapes.multipart}   %
\usetikzlibrary{shapes.geometric}   %
\usetikzlibrary{arrows}
\usetikzlibrary{arrows.meta}        %
\usetikzlibrary{backgrounds}
\usetikzlibrary{patterns}
\usetikzlibrary{fit}

\usepackage{pgfplots}
\usepackage{pgfplotstable}
\usepgfplotslibrary{groupplots}

\newcommand*\Circled[1]{%
	\tikz[baseline=(char.base)]{%
		\node[shape=circle, draw=none, fill=gray!40, thick, inner sep=0.6pt] (char) {%
			\textcolor{black}{\sffamily#1}}; }}

\usepackage{anyfontsize}

\usepackage{amsmath}    %
\usepackage{amsfonts}
\usepackage{amsthm}
\usepackage{amssymb}    %
\usepackage{mathtools}
\usepackage{bm}

\usepackage{caption}
\usepackage{subcaption}
\usepackage[normalem]{ulem}
\usepackage{stackengine}
\usepackage{rotating}	%

\usepackage[scale=.92]{tgheros}

\hypersetup{
	pdftitle={cuSZ-i: High-Ratio Scientific Lossy Compression on GPUs with Optimized Multi-Level Interpolation},
	pdfauthor={Liu, Tian, Wu, et al.},
	pdfsubject={cuSZ-i, camera-ready},
	pdfkeywords={GPU; compression; data quality; lossy; error bound; HPC; scientific data}
}

\usepackage[outercaption]{sidecap}
\sidecaptionvpos{figure}{b}
\sidecaptionvpos{table}{b}

\usepackage[style=ieee, citestyle=numeric, sortcites=true, backend=bibtex, maxbibnames=99]{biblatex}
\usepackage{listings}
\usepackage{lmodern}
\usepackage{microtype}
\microtypesetup{nopatch={footnote}}

\newcommand{\EDIT}[2]{%
    \bgroup%
    \colorbox{B}{\color{white}#1:}\color{B} #2
    \egroup%
}

\newcommand{\FBOXDEBUG}[1]{\setlength{\fboxsep}{0pt}\fbox{#1}}
\renewcommand{\FBOXDEBUG}[1]{#1}

\pgfplotsset{compat=1.15} 

\usepackage[]{titlesec}

\renewcommand\thesubsubsection{\Alph{subsection}.\arabic{subsubsection}}
\renewcommand\theparagraph{\underline{\Alph{subsection}.\arabic{subsubsection}.\alph{paragraph}}}

\titleformat{\subsubsection}[runin]
{\normalfont\normalsize\itshape}{\scalebox{.8}{$\blacksquare$} \thesubsubsection)}{.6em}{}[\ \ ]
\titlespacing*{\subsubsection}
{0pt}{.3ex plus .2ex minus .2ex}{.3ex plus .2ex minus .2ex}

\titleformat{\paragraph}[runin]
{\normalfont\normalsize\itshape}{\theparagraph}{.5em}{}[\ \ ]
\titlespacing*{\paragraph}
{0pt}{.2ex plus 1ex minus .2ex}{.2ex plus .2ex}

\newcommand{\SQUARE}{\scalebox{.8}{$\square$}}

\begin{document}

\title{\thiswork: High-Ratio Scientific Lossy Compression on GPUs with Optimized Multi-Level Interpolation}

\newcommand{\EqContribMark}{$^{\bullet}$}
\newcommand{\UcrMark}{\IEEEauthorrefmark{1}}
\newcommand{\IuMark}{\IEEEauthorrefmark{2}}
\newcommand{\AnlMark}{\IEEEauthorrefmark{3}}
\newcommand{\FsuMark}{\IEEEauthorrefmark{4}}
\newcommand{\UiowaMark}{\IEEEauthorrefmark{5}}
\newcommand{\CasMark}{\IEEEauthorrefmark{6}}
\newcommand{\UhustonMark}{\IEEEauthorrefmark{7}}

\newcommand{\CorrespondingAuthorMark}{$^{\star}$}

\renewcommand{\EqContribMark}{$^{\bullet}$}
\renewcommand{\UcrMark}{$^\triangleright$}
\renewcommand{\IuMark}{$^\circ$}
\renewcommand{\AnlMark}{$^\diamond$}
\renewcommand{\FsuMark}{$^\P$}
\renewcommand{\UiowaMark}{$^\S$}
\renewcommand{\CasMark}{$^\|$}
\renewcommand{\UhustonMark}{$^\times$}

\author
{\normalsize\null\\[-4.2ex] %
\begin{minipage}{\linewidth}\centering
Jinyang Liu\,\UhustonMark\EqContribMark,
Jiannan Tian\,\IuMark\EqContribMark,
Shixun Wu\,\UcrMark\EqContribMark,%
    \thanks{\hspace{-1em}\EqContribMark\,Jinyang Liu, Jiannan Tian, and Shixun Wu contributed equally to this work.}
Sheng Di\,\AnlMark\CorrespondingAuthorMark,
Boyuan Zhang\,\IuMark,
Robert Underwood\,\AnlMark,
Yafan Huang\,\UiowaMark,\\
Jiajun Huang\,\UcrMark,
Kai Zhao\,\FsuMark,
Guanpeng Li\,\UiowaMark,
Dingwen Tao\,\IuMark\CorrespondingAuthorMark, 
    \thanks{\hspace{-1em}\CorrespondingAuthorMark\,Corresponding authors: Sheng Di, Dingwen Tao.}
Zizhong Chen\,\UcrMark,
Franck Cappello\,\AnlMark
\\[.9ex]
\UhustonMark\,%
University of Houston, Houston, TX, USA;\ \ \texttt{jliu217@central.uh.edu}\\
\IuMark\,%
Indiana University, Bloomington, IN, USA;\ \ \texttt{\,\string{\,jti1,\,bozhan,\,ditao\,\string}@iu.edu}\\
\UcrMark\,%
University of California, Riverside, CA, USA;\ \ \texttt{\string{\,swu264,\,jhuan380\,\string}@ucr.edu, chen@cs.ucr.edu}\\
\AnlMark\,%
Argonne National Laboratory, Lemont, IL, USA;\ \ \texttt{\string{\,sdi1,\,runderwood\,\string}@anl.gov, cappello@mcs.anl.gov}\\
\UiowaMark\,%
University of Iowa, Iowa City, IA, USA;\ \ \texttt{\string{\,yafan-huang,\,guanpeng-li\,\string}@uiowa.edu}\\
\FsuMark\,%
Florida State University, Tallahassee, FL, USA;\ \ \texttt{kai.zhao@fsu.edu}%
\end{minipage}\\[-3ex]
}

\thispagestyle{plain}\pagestyle{plain}

\maketitle

\begin{abstract}

	Error-bounded lossy compression is a critical technique for significantly reducing scientific data volumes. Compared to CPU-based compressors, GPU-based compressors exhibit substantially higher throughputs, fitting better for today's HPC applications. However, the critical limitations of existing GPU-based compressors are their low compression ratios and qualities, severely restricting their applicability.
	To overcome these, we introduce a new GPU-based error-bounded scientific lossy compressor named {\thiswork}, with the following contributions:
	(1) A novel GPU-optimized interpolation-based prediction method significantly improves the compression ratio and decompression data quality.
	(2) The Huffman encoding module in {\thiswork} is optimized for better efficiency.
	(3) {\thiswork} is the first to integrate the NVIDIA Bitcomp-lossless as an additional compression-ratio-enhancing module.
	Evaluations show that {\thiswork} significantly outperforms other latest GPU-based lossy compressors in compression ratio under the same error bound (hence, the desired quality), showcasing a 476\% advantage over the second-best.
	This leads to {\thiswork}'s optimized performance in several real-world use cases.

\end{abstract}

\section{Introduction}\label{cuszinterp::intro}

Large-scale scientific applications and advanced experimental instruments produce vast data for post-analysis, creating exascale
scientific databases in supercomputing clusters. For instance, Hardware/Hybrid Accelerated Cosmology Code (HACC)~\cite{hacc,miraio} may produce petabytes of data over hundreds of snapshots when simulating 1 trillion particles. Those extremely large databases raise tough challenges for the management and utility of data. To this end, data reduction is becoming an effective method to resolve this big data issue. Although traditional methods of lossless data reduction can guarantee zero information loss, they suffer from limited compression ratios.
Specifically, lossless compression by roughly 2:1 to 3:1~\cite{son2014data, zhao2021optimizing}. Over the past years, error-bounded lossy compressors have strived to address the high compression ratio requirements for scientific data: they get not only very high compression ratios~\cite{liang2018error,sz3,qoz,SPERR} but also perform strict control over the data distortion regarding various modes of user-set error bounds for post-analysis. Notably, CPU-based lossy compressors have assisted scientific simulation to achieve longer simulating time (e.g., CESM-LE~\cite{use-case-Franck}), and larger scales (e.g., quantum circuit simulation~\cite{shengdi-quantum-sim-2019}).

    {Yet, there are scenarios where even state-of-the-art CPU-based scientific compressors fall short and throttle performance. \Circled{1} Current large-scale scientific simulations have offloaded the computationally intensive components to GPU \cite{hacc,nyx,rtm,qmcpack}~such that GPU, with its limited space, resides large amounts of data temporarily, and moving them to the main memory incurs undesired latency.
        \Circled{2} Advanced instruments have unprecedented peak data acquisition rates, e.g., LCLS-II~\cite{lcls} X-ray imaging can top at 1 TB/s~\cite{underwood2023roibin} (\citeyear{underwood2023roibin}), far beyond what CPU-based compressors can handle.} As a reference, QoZ~\cite{qoz}~\footnote{QoZ, as an interpolation-based scientific error-bounded lossy compressor, is the state-of-the-art CPU-SZ variant in terms of rate-distortion.} achieves a single-core compressing rate of up to 0.23 GB/s.
Thus, high-throughput GPU-based scientific lossy compressors have been developed for in situ data compression, such as {\cusz}~\cite{cusz2020,cuszplus2021}, cuSZx~\cite{szx}, FZ-GPU~\cite{FZGPU}, cuSZp~\cite{cuszp}, and cuZFP~\cite{cuZFP}, topping at tens to hundreds of gigabytes per second compression throughputs per GPU. Nevertheless, they suffer from a low compression ratio (or quality) at each quality (or ratio) constraint.
For instance, although {\cusz} has achieved the highest compression ratio among existing GPU-based compressors, it typically only reaches about 10\% to 30\% (data-dependent) of the CPU-based SZ3 compressor at the same PSNR.
Thus, existing works have not fulfilled the requirement of high-ratio-quality error-bounded lossy compression for scientific data on GPU platforms.

In this work, we leverage the strengths of the existing CPU- and GPU-based work to design the new GPU-based error-bounded lossy compressor.
Specifically, we expect to achieve sufficient compression throughput, far over typical CPU-based ones, and adequate compression ratios, which are improved over the mentioned GPU-based ones.
The challenges are three-fold: \textit{first}, many CPU-prototyped predictors carry intrinsic data dependency, inhibiting parallelization; \textit{second}, sophisticated CPU-side tuning techniques that are proven effective can incur high latency even considering customization for GPU; \textit{third}, most high-ratio lossless encoders in existing error-bounded lossy compressors only show poor throughputs on GPU platforms.
Responding to these challenges, we introduce a novel, fine-tuned, highly parallelized interpolation-based data predictor. We also introduce a synergetic lossless scheme by coupling the improved Huffman encoding and additional de-redundancy pass and demonstrate it using NVIDIA {\bitcomp}.
Having overcome the challenges above,
we propose a GPU-based error-bounded lossy compressor with both high compression throughput and effectiveness, named {\thiswork}.
Overall, {\thiswork} preserves a high compression throughput within the same magnitude as existing GPU-based compressors and profoundly prevails over all others regarding compression ratio and data quality. To our best knowledge, {\thiswork} is the first and the only high-ratio and high-quality GPU-based scientific error-bounded lossy compressor, which fills the major gap between the compression ratio and quality of CPU-based and GPU-based scientific lossy compressors.

We summarize our contributions as follows:

\begin{itemize}[leftmargin=1.3em]
    \item We develop a GPU-optimized interpolation-based data predictor {\ginterp} with highly parallelized efficient interpolation, which can present excellent data prediction accuracy, thus leading to a high overall compression ratio.
    \item We design and implement a lightweight interpolation auto-tuning kernel for GPU interpolation to optimize both the performance and compression quality of {\thiswork}.
    \item We improve the implementation of GPU-based Huffman encoding and import a new lossless module to reduce its encoding redundancy further.
    \item {\thiswork} substantially improves compression ratio over other state-of-the-art GPU-based scientific lossy compressors up to 476\% under the same error bound or PSNR. Meanwhile, it preserves a compression throughput of the same magnitude as other GPU compressors.
\end{itemize}

The rest of this paper is arranged as follows: \SEC\ref{cuszinterp::related} introduces related works. \SEC\ref{cuszinterp::background} provides the background and motivation for our research. \SEC\ref{cuszinterp::design} demonstrates the framework of {\thiswork}. The design of {\thiswork} interpolation-based predictor: {\ginterp} is illustrated in detail in \SEC\ref{cuszinterp::design::predictor}, and the new design of {\thiswork} lossless encoding modules is proposed in \SEC\ref{cuszinterp::design::lossless}. In \SEC\ref{cuszinterp::eval}, the evaluation results are presented and analyzed. Finally, \SEC\ref{cuszinterp::conclusion} concludes our work and discusses future work.

\section{Related Work}
\label{cuszinterp::related}

Scientific data compression has been studied for years to address storage burdens and I/O overheads. The compression techniques are in two classes: lossless and lossy. Compared to lossless compression, lossy compression can provide a much higher compression ratio at the cost of information loss. Moreover, scientific computing practice often requires the error to be quantitatively determined for accurate post-analysis.

Recently, many error-bounded lossy compressors for scientific data have been developed, such as SZ~\cite{zhao2021optimizing,sz3,hybrid-sz}, ZFP~\cite{zfp}, QoZ~\cite{qoz}, SPERR~\cite{SPERR}, TTHRESH~\cite{ballester2019tthresh}, and several AI-assisted works~\cite{ae-sz,SRN-SZ,han2022coordnet}. The usability of these compressors lies in allowing users to strictly control accuracy loss in data reconstruction and post-analysis across various domains.
these specialized compressors are also crucial for the treatment of scientific data, one important aspect of which is high-dimensional data interpretation. Previous work~\cite{tac2022,amric2023,tacplus2024}
emphasize the high-dimensional data continuity, necessitating the preservation of such information. To this end, we are set to design an efficient high-dimensional data compressor.

Moreover, considering the increasing data generation speed of scientific instruments/simulations, and the popularity of heterogenous HPC systems and applications that directly yield data on GPU platforms, GPU-based scientific lossy compressors have been proposed to deliver orders of magnitude higher throughputs than CPU compressors. Specifically, cuZFP~\cite{cuZFP}, the CUDA implementation of transform-based ZFP, leverages discrete orthogonal transform and embedding encoding for compression, allowing the user to specify the desired bit rate. Also, prediction-based GPU compressors have been proposed, including {\cusz}, cuSZx~\cite{szx}, FZ-GPU~\cite{FZGPU}, and cuSZp~\cite{cuszp}. While sharing the error-boundness in the reconstructed data, these prediction-based compressors can fit into different use scenarios, as detailed below.
1) {\cusz}~\cite{cusz2020,cuszplus2021} features fully parallelized prediction-quantization with outlier compacted and coarse-grained Huffman coding that encodes multibyte quantization codes using bits inverse to their frequencies;
2) cuSZx~\cite{szx} features a monolithic design to deliver extremely high throughput at the cost of lower data quality and compression ratio.
3) FZ-GPU \cite{FZGPU} alters the compression pipeline by replacing the entire lossless encoding stage with bit-shuffle and dictionary encoding, aiming to deliver higher throughput.
4) cuSZp~\cite{cuszp} modifies {\cusz} by fusing prediction-quantization and 1D blockwise encoding subroutine into a monolithic GPU kernel, practically achieving high end-to-end throughput.
All the high-throughput-oriented GPU-based compressors share the limitation of suboptimal compression ratios.
For example, {\cusz}, FZ-GPU, and cuSZp all feature Lorenzo prediction, resulting in the ceiled data quality, as described in~\cite{zhao2021optimizing}. Furthermore, note that \mbox{MGARD-X} (CUDA-backend)~\cite{repo-mgardx} is another GPU compressor practice; though it achieves comparable rate-distortion to {\cusz}, it has been reported to be significantly lower in throughput than other GPU-based compressors~\cite{FZGPU}, making it not fit our use scenarios. Therefore, we exclude it from the discussion.

\section{Background and Research Motivation}
\label{cuszinterp::background}

In this section, we elaborate on our research background.
First, we demonstrate {\cusz} (the CUDA version of SZ)~\cite{cusz2020,cuszplus2021} because its prediction-based nature typically results in a higher compression ratio and makes it the fundamental of \thiswork.
Next, we specify our research motivation by analyzing the characteristics and limitations of existing GPU-based scientific lossy compressors, then set up our research target.

\subsection{{\cusz} Framework}\label{cuszinterp::background::cusz}

\cusz~\cite{cusz2020,cuszplus2021} is a GPU-based scientific error-bounded lossy compression, which shows state-of-the-art compression ratios and distortions among GPU-based error-bounded scientific lossy compressors.
Like the CPU SZ framework, {\cusz} also has a modular compression framework composed of data prediction, data quantization, and lossless encoding.
In these frameworks, predictions are made for each element of the input data, and the predicted values adjusted by {\quantcode}s (denoted by $q$'s) are accurate with respect to the user-specified error bound and replace the original data values. The {\quantcode}s are either encoded when $|q| < R$, an internal parameter, or otherwise saved as \textit{outliers} when they are too big (i.e., $|q|\ge R$) for efficient encoding.
    {\cusz} features fully parallelized Lorenzo prediction-quantization kernels for compression and decompression. The lossless encoding component features a GPU-based coarse-grained parallelized Huffman encoding, which is the source of the relatively high compression ratios. {\cusz} crucially differs from CPU SZ as it does not have a further pass of de-redundancy lossless encoding (e.g., Zstd). This is considered a tradeoff between the throughput and the compression ratio because the sophisticated logic in the redundancy-canceling component can significantly deteriorate the data-processing performance. We refer readers to the {\cusz} papers~\cite{cusz2020,cuszp} for more details.

\subsection{Research Motivations}
In this part, we discuss several key limitations of existing GPU-based scientific lossy compressors such as {\cusz}, then propose our design target of  {\thiswork} to address those issues.
\subsubsection{Limitation of existing GPU-based lossy compressors in data reconstruction quality}
The first limitation of existing GPU-based lossy compressors is that their data reconstruction quality is sub-optimal in various cases. As mentioned before in \SEC~\ref{cuszinterp::related}, the Lorenzo data predictor leveraged by {\cusz}, cuSZp, and cuSZx have been proven to be inaccurate in many cases \cite{zhao2021optimizing}, so their compression ratios at the desired data qualities are far lower than interpolation-based scientific compressors on CPU such as SZ3~\cite{zhao2021optimizing,sz3}. Therefore, we would like to learn from the advanced predictor design of high-ratio CPU data compressors and propose a GPU-customized high-accuracy data predictor for error-bounded lossy compression on GPUs.

\subsubsection{Limitation of existing GPU-based lossy compressors due to lossless encoding modules}
Another critical limitation of existing GPU-based lossy compressors is that their lossless encoding modules produce smaller compression ratios due to the concern of compression throughput. Taking {\cusz} as an example, its sole lossless encoding module, Huffman encoding, uses at least 1 bit to map each data element, making the compression bit rate always higher than 1. More importantly, the compressed data of {\cusz} still has high redundancy in broad cases. Therefore, we need to address this issue by designing or employing new lossless encoding techniques.

\subsubsection{Design goals of {\thiswork}}

Overall, we endeavor to significantly boost the compression quality of existing GPU-based error-bounded lossy compressors by applying a series of optimizations to address the above issues. Our development of {\thiswork} contains and is not limited to \Circled{1} leveraging a more effective data prediction scheme for better compression quality, and \Circled{2} integrating high-performance lossless modules in the pipeline for better compression ratios. As noted, there are three primary challenges. \textit{First},
the existing CPU-based high-accuracy data predictor, e.g., SZ3 and QoZ interpolators, features many levels of interpolations and will result in heavy data dependency and low throughputs if directly ported to GPUs.
\textit{Second}, we need to effectively configure the predictor to boost the compression ratio on GPU. \textit{Third}, we need to efficiently utilize the GPU resource for usable throughputs in real-use scenarios.
In the following \SEC\ref{cuszinterp::design}, we will present how we overcame this challenge by proposing a novel GPU-optimized design of the data predictor to optimize the prediction throughput on GPUs.

\begin{SCfigure*}[1.0][!htbp]
    \FloatBodyStyle\footnotesize
    \begin{tikzpicture}[]

    \tikzset{
        every node/.style={align=center, draw, rounded corners=2pt, font=\footnotesize, node distance=.2in, fill=white},
        >=latex,
        label/.style={align=center, draw=none},
        diff/.style={left color=B!30!white, right color=B!10!white},
    };

    \coordinate (ori) at (0, 0);

    \node[below=.22in of ori, draw=none] (cusz) {\makebox[3em][l]{\cusz}};
    \node[below=.3in of cusz, draw=none] (ours) {\makebox[3em][l]{\thiswork}};
    
    \node[draw=none, align=left, right=.1in of ours, fill=none] (ours-pred-0) {\textbf{\S\ref{cuszinterp::design::predictor}} \textbf{spline} \\ \textbf{interpolation}\\+ outlier filter};
    \node[fill=white, draw=RoyalBlue!50, inner sep=2pt, right=.0in of ours-pred-0.north east, anchor=north west] (ours-pred-1) {\scriptsize \S\ref{cuszinterp::design::interp-config} \ginterp};
    \node[fill=white, draw=RoyalBlue!50, inner sep=2pt, right=.0in of ours-pred-0.east, anchor=west] (ours-pred-2) {\scriptsize \S\ref{cuszinterp::autotuning} auto-tuning};
    \node[fill=white, draw=RoyalBlue!50, inner sep=2pt, right=.0in of ours-pred-0.south east, anchor=south west] (ours-pred-2) {\scriptsize \S\ref{cuszinterp::design::par-spline3} parallel algo.};

    \begin{scope}[on background layer]
        \node[diff, fit=(ours-pred-0) (ours-pred-1) (ours-pred-2)] (ours-pred) {};
    \end{scope}
    \node[right=of ours-pred] (ours-hist) {histogram};
    \node[right=of ours-hist] (ours-hf) {multibyte\\Huffman enc.};
    \node[diff, right=of ours-hf] (ours-bc) {de-redundancy\\enc. (e.g., Bitcomp)};

    \node[] (cusz-pred) at (cusz -| ours-pred) {dual-quant \textbf{Lorenzo} \\+ outlier filter};
    \node[] (cusz-hist) at (cusz -| ours-hist) {histogram};
    \node[] (cusz-hf) at (cusz -| ours-hf) {multibyte\\Huffman enc.};

    \node[label] at (ori -| ours-pred) {\scshape prediction-quantization};
    \node[label] at (ori -| ours-hist) {\scshape statistics};
    \node[label] at (ori -| ours-hf) {\scshape lossless enc. (1)};
    \node[label] at (ori -| ours-bc) {\scshape lossless enc. (2)};
    
    \begin{scope}[on background layer]
        \coordinate (tmp-bottom) at ([yshift=-.4em]ours-hf.south);
        \node[fit=(tmp-bottom) (cusz-hf), rounded corners=4pt, draw=lightgray, very thin, fill=lightgray, fill opacity=.2] (tweak) {};
        \node[right=0pt of tweak.north east, anchor=north west, draw=none] {\scriptsize \S\ref{cuszinterp::design::lossless::tweak} tweak};
        	
        \coordinate (tmp-left) at ([xshift=-.4em]ours-hf.west);
        \node[fit=(tmp-left) (ours-bc), rounded corners=4pt, draw=lightgray, very thin, fill=lightgray, fill opacity=.2] (synergy) {};
        \node[right=0pt of synergy.north east, anchor=south east, draw=none] {\scriptsize \S\ref{cuszinterp::reason-why-bitcomp} encoding synergy};
    \end{scope}

    \draw[->] (cusz-pred) -- (cusz-hist);
    \draw[->] (cusz-hist) -- (cusz-hf);
    \draw[->] (ours-pred) -- (ours-hist);
    \draw[->] (ours-hist) -- (ours-hf);
    \draw[->] (ours-hf) -- (ours-bc);

\end{tikzpicture}

    \caption{The compression pipe\-line of {\thiswork} and the comparison between it and {\cusz} one. The shading indicates the differentiators of {\thiswork} from one design basis, {\cusz}.\vspace{-4ex}}
    \label{cuszinterp::fig::overview}
    \vspace{-4ex}
\end{SCfigure*}
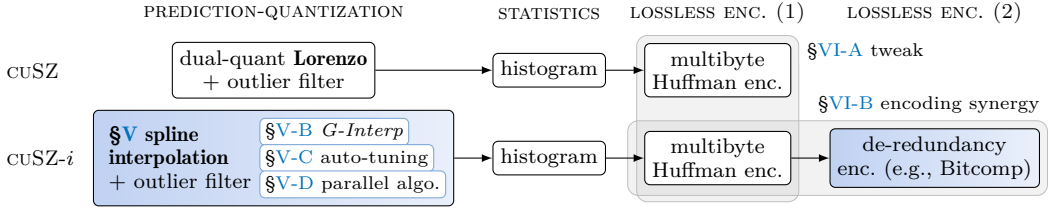

\section{{\thiswork} Design Overview}\label{cuszinterp::design}

In this section, we present the design of {\thiswork}.
First,
\FIG\ref{cuszinterp::fig::overview} presents the pipelines of {\cusz} and {\thiswork}.
Like {\cusz} and many other GPU-based compressors, {\thiswork} is also a prediction-based error-bounded lossy compressor that features data prediction, error quantization, and lossless encoding. In the compression pipeline, {\thiswork} follows the CPU-SZ/{\cusz} to approximate input data with a data predictor, and the prediction errors are quantized and recorded.
The quantized errors are further encoded and stored to be used in the decompression process. The core innovations in {\thiswork} are as follows.
\begin{itemize}[leftmargin=1.3em]
    \item We develop a practically GPU-optimized interpolation-based data predictor with efficient parallelization to replace the Lorenzo data predictor~\cite{sz17,liang2018error}, leading to much higher data prediction accuracy and better rate-distortion.
    \item We integrate a lightweight profiling-based auto-tuning mechanism into its kernel to jointly optimize the quality and performance of the interpolation-based data predictor.
    \item To enhance the throughput when Huffman-encoding {\quantcode}s, the component of building the histogram is optimized.
    \item To maximally boost the compression ratio with minor speed degradation, we \emph{can} enable another lossless encoding pass (e.g., {\bitcomp}) after Huffman encoding in \thiswork.
\end{itemize}

In the remainder of this section, we will detail the innovations we used to design the interpolation-based data predictor practically optimized on GPU.

\section{{\thiswork} Interpolation-based Data Predictor}\label{cuszinterp::design::predictor}

Based on CPU interpolation-based data predictor design~\cite{zhao2021optimizing,qoz,hpez}, {\thiswork} introduces a new interpolation-based prediction scheme, named {\ginterp}, for GPU platforms featuring hardware-software codesign.
The new scheme consists of a spline interpolative predictor utilizing anchor points and an auto-tuning strategy. As we will show in most use scenarios, the new interpolative predictor has advantages over the Lorenzo predictor in terms of prediction accuracy and compression ratio.

\subsection{Basic Design Concept of {\ginterp}}\label{cuszinterp::design::predictor::ginterp}

\begin{figure}[!htb]
    \includegraphics[width=\linewidth]{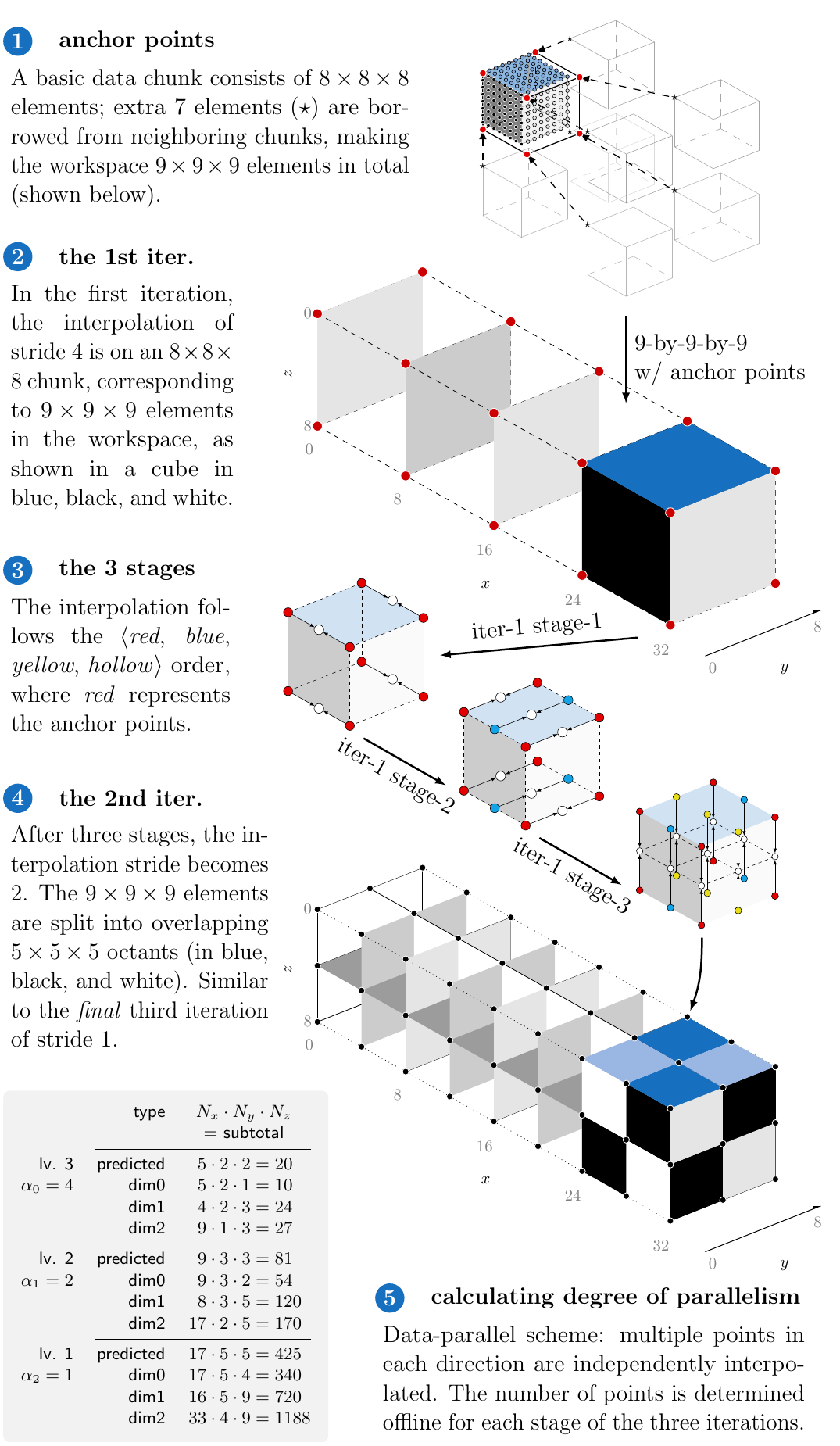}
    \caption{{\ginterp}, the parallelized interpolation-based data predictor.}
    \label{cuszinterp::fig::data-parallel}
\end{figure}

\FIG\ref{cuszinterp::fig::data-parallel} demonstrates the basic design of {\ginterp} with a 3D example.
Though accurate, the existing CPU-based interpolation data predictors~\cite{zhao2021optimizing,qoz,hpez} are slow (e.g., 0.23 GB/s) and cannot directly be ported to GPU platforms. {\ginterp} becomes a re-design according to the hardware trait to maximize its data-processing parallelism. Compared to the CPU-based interpolation, it exhibits an accuracy-parallelism tradeoff but still preserves substantially higher data prediction accuracy than the baseline Lorenzo.
To these ends,
we partition the input data into relatively small chunks and eliminate the data dependencies across them.
In the illustration, the 3D input data is partitioned into $8^3$ chunks (or $16^2$ for 2D chunks/512 for 1D chunks).
Next, to avoid cross-chunk data dependencies, the interpolation operations need to be confined to limited units of data chunks within a short range.
Inspired by QoZ~\cite{qoz}, we introduced anchor points in {\ginterp} that are losslessly stored so that all interpolations are bounded in the range of any 2 adjacent anchor points (i.e., within one single data chunk). For example, in the 3D case shown in \FIG\ref{cuszinterp::fig::data-parallel}-\Circled{1}, in each data chunk, one vertex point is assigned as an anchor point (red points in \FIG\ref{cuszinterp::fig::data-parallel}), and 7 other anchor points (on the rest vertices) are borrowed from the surrounding chunks. Afterward, the interpolation within each data chunk is parallelizable and independent of others. In the 3D input data array, approximately 1 of 512 elements becomes anchor points to preserve,
whose saving overhead can be further decreased by the additional pass of lossless encoding.

Once the partitioned data chunks and anchor points are set up, as shown in \FIG\ref{cuszinterp::fig::data-parallel}-(\Circled{3}--\Circled{4}), the interpolation-based data predictions in each chunk are performed in parallel.
Like in existing works~\cite{zhao2021optimizing,qoz}, the interpolations are performed level by level, from a large stride to smaller ones (from 4 to 2 and 1 for {\ginterp}), and each chunk is expanded from 
$1^3$ to $2^3$, $4^3$, lastly $8^3$. 
At each level, interpolations are performed subsequently along each dimension. Moreover, within each chunk, the interpolations on the same level and along the same dimension are also performed in parallel without dependencies. Later, we will feature essential details of this process, including the interpolation splines, error bounds, and the order of interpolation direction.

\subsection{Interpolation Configurations of {\ginterp}}\label{cuszinterp::design::interp-config}

\subsubsection{Interpolation splines} \label{cuszinterp::design::spline-types}

\begin{figure}[ht]
    \centering
    \includegraphics[width=\linewidth]{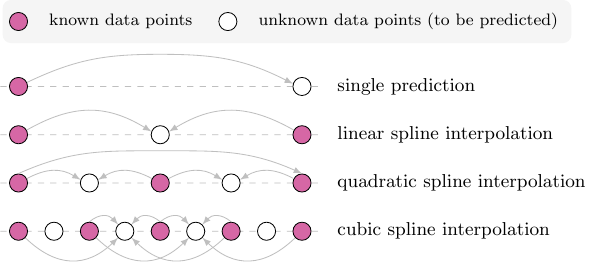}
    \caption{Interpolation splines.}
    \label{cuszinterp::fig::splines}
    \vspace{-4ex}
\end{figure}

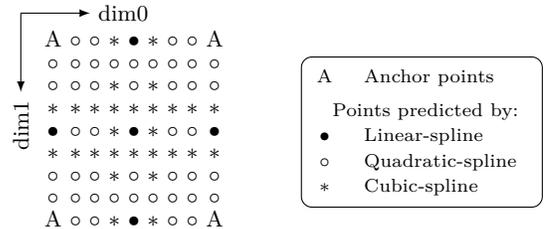
\begin{figure}[ht]
    \FloatBodyStyle
    \begin{tikzpicture}
  \node (demo) {
    \renewcommand{\arraystretch}{.85}
    \setlength{\tabcolsep}{1.5pt}
    \begin{tabular}{ @{} *{9}{c} @{}}
      A         & $\circ$ & $\circ$ & $*$     & $\bullet$ & $*$     & $\circ$ & $\circ$ & A         \\
      $\circ$   & $\circ$ & $\circ$ & $\circ$ & $\circ$   & $\circ$ & $\circ$ & $\circ$ & $\circ$   \\
      $\circ$   & $\circ$ & $\circ$ & $*$     & $\circ$   & $*$     & $\circ$ & $\circ$ & $\circ$   \\
      $*$       & $*$     & $*$     & $*$     & $*$       & $*$     & $*$     & $*$     & $*$       \\
      $\bullet$ & $\circ$ & $\circ$ & $*$     & $\bullet$ & $*$     & $\circ$ & $\circ$ & $\bullet$ \\
      $*$       & $*$     & $*$     & $*$     & $*$       & $*$     & $*$     & $*$     & $*$       \\
      $\circ$   & $\circ$ & $\circ$ & $*$     & $\circ$   & $*$     & $\circ$ & $\circ$ & $\circ$   \\
      $\circ$   & $\circ$ & $\circ$ & $\circ$ & $\circ$   & $\circ$ & $\circ$ & $\circ$ & $\circ$   \\
      A         & $\circ$ & $\circ$ & $*$     & $\bullet$ & $*$     & $\circ$ & $\circ$ & A         \\
    \end{tabular}%
  };

  \draw[-latex] ([yshift=1ex]demo.north west) -- ([xshift=3em, yshift=1ex]demo.north west) node [anchor=west] {dim0};
  \draw[-latex] ([yshift=1ex]demo.north west) -- ([yshift=-3em]demo.north west) node[anchor=east, rotate=90] {dim1};

  \node [rounded corners, draw, right=3em of demo] {
    \renewcommand{\arraystretch}{1.2}
    \scriptsize
    \begin{tabular}{@{}cl@{}}
      A         & Anchor points                \\[1ex]
      \multicolumn{2}{l}{Points predicted by:} \\
      $\bullet$ & Linear-spline                \\
      $\circ$   & Quadratic-spline             \\
      $*$       & Cubic-spline
    \end{tabular}

  };
\end{tikzpicture}
    \caption{Illustration of spline interpolations using a 2D slice.%
    }
    \label{cuszinterp::fig::spline-on-2D-slice}
    \vspace{-1ex}
\end{figure}

Shown in \FIG\ref{cuszinterp::fig::data-parallel}, for the compression and decompression of a multi-dimensional data grid, {\ginterp} performs the interpolations with 1D splines, executed along different dimensions. For the specific 1D splines, \FIG\ref{cuszinterp::fig::splines} presents an example with 1D data slices. For each interpolation (to predict $x_n$ with a predicted value $p_n$), depending on the number of available neighbor points, there are four circumstances:
\begin{itemize}[leftmargin=1.3em]
    \item 4 neighbors available: The cubic spline interpolation with $x_{n-3}$, $x_{n-1}$, $x_{n+1}$, and $x_{n+3}$ is computed to predict $x_n$;
    \item 3 neighbors available: The quadratic spline interpolation with $x_{n-3}$, $x_{n-1}$,
          and $x_{n+1}$ (or $x_{n-1}$, $x_{n+1}$, and $x_{n+3}$) is computed to predict $x_n$;
    \item 2 neighbors available: The linear spline interpolation with $x_{n-1}$ and $x_{n+1}$ is computed to predict $x_n$;
    \item 1 neighbor available: $x_{n-1}$ serves as a prediction of $x_n$.
\end{itemize}
As mentioned, because the predicted values with {\quantcode} adjustment replace the original value, the decompression precisely replays the interpolation-based prediction from the {\quantcode}. %
\FIG\ref{cuszinterp::fig::spline-on-2D-slice} demonstrates the interpolation on a $9\times9$ 2D grid with an anchor stride of 8. The predicted data points with different interpolation splines are marked with their corresponding shapes.
We list each spline function in {\ginterp} as follows:
\begin{itemize}[leftmargin=1.3em]
    \item Linear Spline:
          \[\textstyle
              p_n= \frac{1}{2}x_{n-1} + \frac{1}{2}x_{n+1}
          \]
    \item Quadratic Spline:
          \[
              \begin{split}
                  p_n               & = \textstyle -\frac{1}{8}x_{n-3} + \frac{6}{8}x_{n-1} + \frac{3}{8}x_{n+1}          \\
                  \text{or\ \ } p_n & = \textstyle \phantom{-}\frac{3}{8}x_{n-1} + \frac{6}{8}x_{n+1} -\frac{1}{8}x_{n+3}
              \end{split}
          \]
    \item Cubic Spline (not-a-knot):
          \[\textstyle
              p_n=-\frac{1}{16}x_{n-3} + \frac{9}{16}x_{n-1} + \frac{9}{16}x_{n+1} -\frac{1}{16}x_{n+3}\]
    \item Cubic Spline (natural):
          \[\textstyle
              p_n=-\frac{3}{40}x_{n-3} + \frac{23}{40}x_{n-1} + \frac{23}{40}x_{n+1} -\frac{3}{40}x_{n+3}\]
\end{itemize}
We refer the reader to \cite{zhao2021optimizing} and \cite{hpez}, from which we derived the function from their analyses.
It is worth noting that two different cubic splines serve the same circumstance because each can outperform the others on different datasets. We will use an auto-tuning model of {\ginterp}  to select a best-fit cubic spline during each compression task (to be explained in \SEC\ref{cuszinterp::autotuning}).

\subsubsection{Level-wise interpolation error bound}
The interpolation-based prediction features the synchronous data prediction and error control process. For each input data point, {\ginterp} quantizes the error of its interpolation-based prediction and adjusts the prediction value with an offset of the quantized error. The adjusted prediction will serve as the decompression output and be used to predict subsequent input data points. Therefore, the prediction quality of early interpolations will impact later ones. As verified by \cite{qoz}, for this prediction paradigm, applying lower error bounds on higher interpolation levels (i.e., with larger interpolation strides) can significantly decrease the compression-introduced distortion with little compression ratio degradation. In some cases, it can even improve both compression ratio and distortion. Thus, similarly with \cite{qoz} and \cite{hpez}, {\ginterp} reduces the error bounds for high-level interpolations according to
$\ell=e\cdot(\alpha^{\ell-1})^{-1}$,
where $\alpha \ge 1$, $e$ is the global error bound, $\ell$ is the interpolation level, and $e_\ell$ is the error bound on level $\ell$. $\alpha$ is a parameter of error-bound reduction. In \SEC\ref{cuszinterp::autotuning}, we will discuss how the specific $\alpha$ value is determined.

\subsection{Profiling-based Auto-tuning of {\ginterp} Interpolations}\label{cuszinterp::autotuning}

Because the prediction accuracy of interpolation-based data predictors is highly sensitive to their configurations~\cite{qoz}, auto-tuning modules are often jointly leveraged with the interpolation-based predictors for preserving the data prediction accuracy~\cite{zhao2021optimizing,qoz,liu2023faz,hpez}. However, existing CPU-based auto-tuning strategies cannot be directly transferred to GPU platforms because they will introduce more computational overhead on GPUs. To this end, we leverage a lightweight profiling-and-auto-tuning kernel for {\ginterp}. This kernel comprises two functionalities: data profiling and interpolation auto-tuning.

\subsubsection{Data-profiling}
We profile the input data in 2 steps. First, we compute its value range to acquire both the \emph{absolute} and value-range-based \emph{relative} error bounds (the former divided by the value range equals to the latter). Second, {\ginterp} uniformly samples a small number of data points from the input (e.g., a $4^3$ sub-grid for 3D cases). For each sampled point, {\ginterp} performs two instances of cubic spline interpolations along each dimension (e.g., totaling $2\times3$ tests for 3D cases). The profiling kernel accumulates the interpolation prediction errors separately for each interpolation spline and interpolation dimension.
\subsubsection{Interpolation auto-tuning}
With the profiling information, {\ginterp} determines the interpolation configurations by the following strategy.
First, {\ginterp} computes the error-bound reduction factor $\alpha$ by a piecewise linear function of the value-range-based relative error bound $\epsilon$:
\begin{equation}\small
    \label{cuszinterp::eq:alpha}
    \alpha=A(\epsilon)=\begin{cases}
        2                                                                & {10^{-1} \leq \epsilon}                            \\
        1.75+0.25\cdot\frac{\epsilon-10^{-2}}{10^{-1}-10^{-2}}           & {10^{-2} \leq \epsilon < 10^{-1}}                  \\
        1.5\phantom{0}+0.25\cdot\frac{\epsilon-10^{-3}}{10^{-2}-10^{-3}} & {10^{-3} \leq \epsilon < 10^{-2}}                  \\
        1.25+0.25\cdot\frac{\epsilon-10^{-4}}{10^{-3}-10^{-4}}           & {10^{-4} \leq \epsilon < 10^{-3}}                  \\
        1\phantom{.00}+0.25\cdot\frac{\epsilon-10^{-5}}{10^{-4}-10^{-5}} & {10^{-5} \leq \epsilon < 10^{-4}}                  \\
        1                                                                & {\hphantom{10^{-5} \leq{}} \epsilon \leq 10^{-5} }
    \end{cases}
\end{equation}

\noindent Dynamically, an optimization of $\alpha$ can be based on both the input data and the value of error bound $\epsilon$. To reduce the computational overhead, we empirically apply this effective calculation instead of a more precise optimization for $\alpha$.
Eq.~\ref{cuszinterp::eq:alpha} is established for the reason that from our and \cite{qoz}'s observations, the optimized value of $\alpha$ is relevant to $\epsilon$ and should decrease with the decrease of $\epsilon$.

Lastly, {\ginterp} evaluates the cubic splines and the smoothness of dimensions by profiling interpolation errors. For each dimension, the cubic spline with the lower error will be applied. Moreover, on each level, the interpolation will start from the least smooth dimension (largest in profiled error) and end in the smoothest dimension (smallest in profiled error). This is explained in \cite{zhao2021optimizing}: the earlier processed dimension will have fewer interpolations performed along it, and vice versa. As a result of performing more interpolations along smoother dimensions, the overall data prediction accuracy can be well optimized.

\subsection{Parallelizing {\ginterp} on GPU}\label{cuszinterp::design::par-spline3}

We practically take advantage of the modern GPU architecture to realize the {\ginterp}'s concepts on the GPU platform.
The algorithm-determined data-dependency challenges are as follows: 1) a portion of data loading and storing are non-coalescing,~\footnote{Memory coalescing is achieved when parallel threads access consecutive global memories to minimize transaction times by loading the same amount of data, allowing for the optimal usage of the global memory bandwidth.}
2) at each interpolation level, dependent traversal stages need to be done, and 3) the finer interpolation depends on the coarser interpolation levels.
Our solutions to meet the challenges sufficiently are stated below.

First, we exploit the data-caching capacity paired with the thread numbers for each thread block. A thread block corresponds to four \textit{basic block} of $8^3$ elements (featured in \SEC\ref{cuszinterp::design::predictor::ginterp}). This builds up a $32_x\!\times8_y\!\times8_z$ chunk for a coalescing load (\FIG\ref{cuszinterp::fig::data-parallel}-\Circled{2}),
minimizing the DRAM transaction. After that, borrowed anchor points are loaded from neighbor blocks before interpolation.
\mbox{\FIG\ref{cuszinterp::fig::data-parallel}-\Circled{3}} shows the first-level interpolation (with a stride of $\alpha_0=4$) in the view of a basic block. After the three-stage interpolation in iteration 1 ($\alpha_0=4$), the inwardly finer interpolation with stride $\alpha_1=2$ continues (\FIG\ref{cuszinterp::fig::data-parallel}-\Circled{4}) and forth for even finer interpolation with $\alpha_2 = 1$.

Next, each GPU thread block interpolates within a $33_x\!\times9_y\!\times9_z$ data block, which encloses the $32_x\!\times8_y\!\times8_z$ input data chunk and borrowed anchor points (\FIG\ref{cuszinterp::fig::data-parallel}-\Circled{2}). The interpolations are performed as described in \SEC\ref{cuszinterp::design::predictor::ginterp}, but now more neighbor points can be available as 4 basic blocks can share their data. For each interpolation level, \FIG\ref{cuszinterp::fig::data-parallel}-\Circled{5} enumerates the interpolated points along each dimension (suppose the interpolations are performed from \texttt{dim0} with length 9 to \texttt{dim2} with length 33). In the level 1 interpolation, the number of interpolated points surpasses the thread number limits of each thread block. Thus, we dynamically assign the number of data points to an active thread.

\subsection{Advantages of {\ginterp}: A Quantitative Analysis}\label{cuszinterp::design::ginterp-advantage}

\begin{figure*}[ht]
    \includegraphics[width=\linewidth]{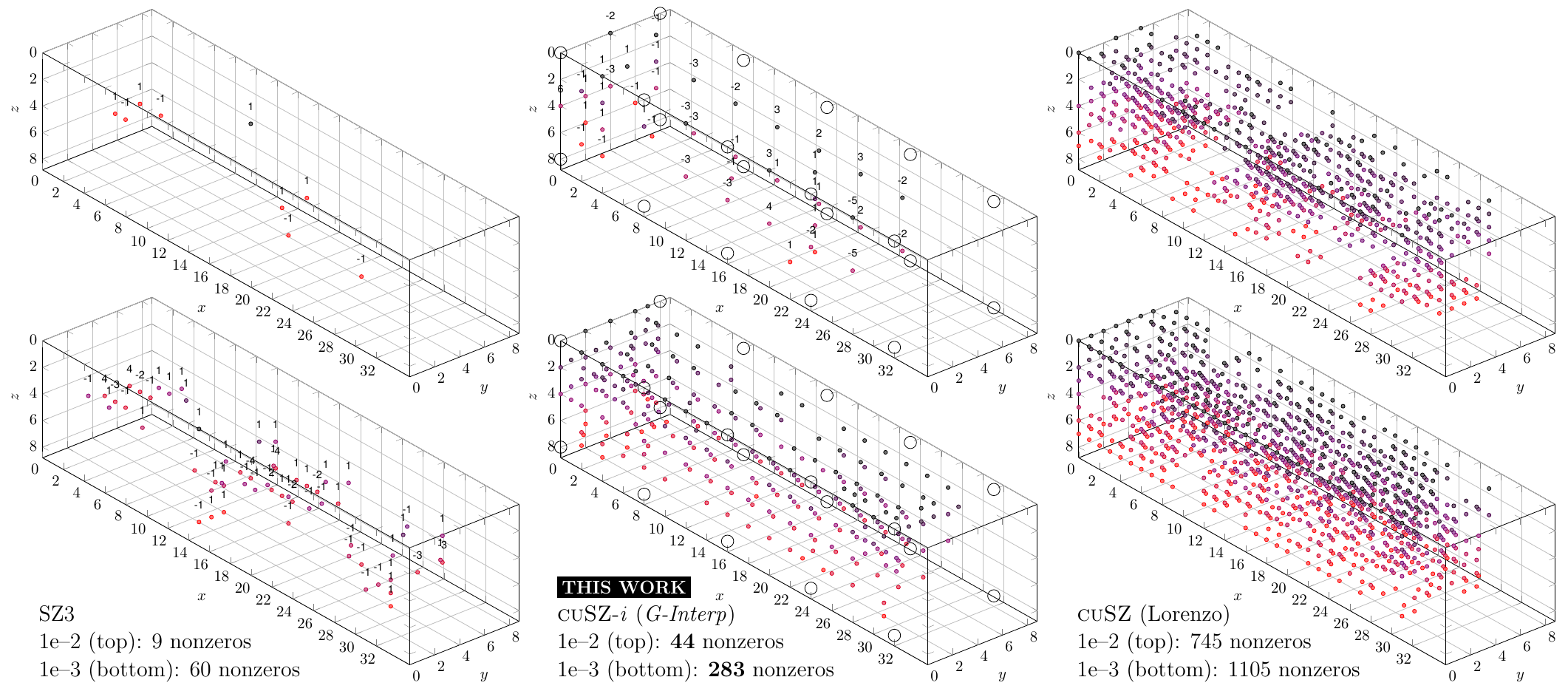}
    \caption{Showcase: counts of nonzero {\quantcode} among CPU SZ3, GPU {\ginterp}, and GPU Lorenzo. Two relative-to-value-range error bounds are used on \texttt{Miranda}-\texttt{Pressure}. Dots in the $33\times9\times9$ bounding box indicate the nonzeros.}
    \label{cuszinterp::fig::showcase-quantcodes}
\end{figure*}

We present two showcases to clearly show how the {\thiswork} {\ginterp} data predictor advantages over Lorenzo predictor in {\cusz}. With better data prediction capability, {\ginterp} generally produces more minor predicting errors. Since those errors will be quantized to a collection of integers, smaller errors will lead to a more concentrated distribution of quantization bins and a higher compression ratio after the Huffman Encoding process.

\begin{figure}[ht]
    \FloatBodyStyle
    \FBOXDEBUG{\includegraphics[width=\linewidth]{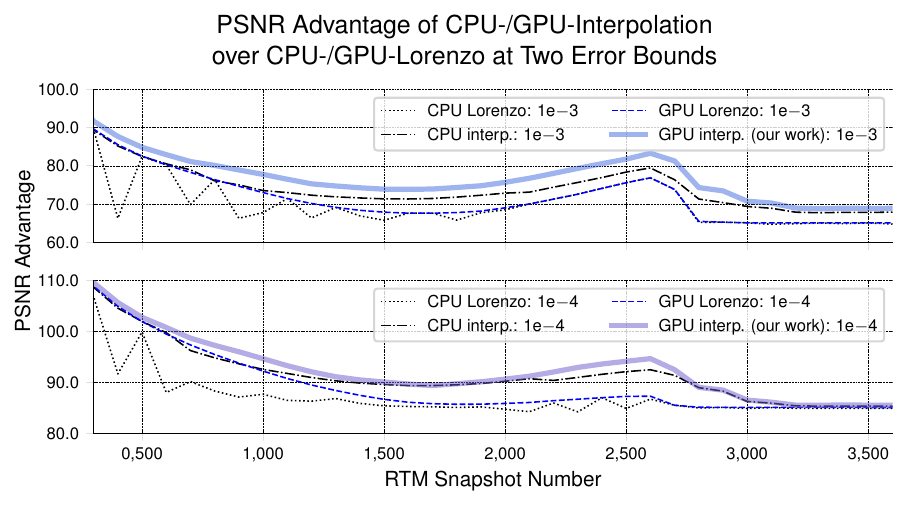}}
    \caption{The PSNR advantage of interpolation over Lorenzo on two error bounds. One snapshot is selected for every 100 among 3700, excluding several ones corresponding to the simulation's initialization phase.}
    \label{fig:rtm-ivl}
\end{figure}

First, \FIG\ref{cuszinterp::fig::showcase-quantcodes} compares the quantized prediction errors of {\cusz}-Lorenzo, {\ginterp}, and CPU-interpolation SZ3 (as a benchmark) when applied to field \texttt{pressure} of \texttt{Miranda} dataset~\cite{miranda}. In the figure, the nonzero {\quantcode}s (i.e., the prediction error is larger than $eb$) are colored according to the amplitude of the {\quantcode}. Under the same error bounds, the {\ginterp} design results in much less nonzero {\quantcode}s than {\cusz}'s Lorenzo predictor and smaller in amplitude, achieving closer outcomes with the CPU-based SZ3.
Next, because {\ginterp} exhibits significantly better data prediction accuracy than {\cusz}-Lorenzo, its data reconstruction fidelity is also substantially improved over the Lorenzo predictor. In \FIG\ref{fig:rtm-ivl}, we propose the comparison between decompression PSNR of {\ginterp} and {\cusz}-Lorenzo (under the same error bound, both CPU-based and GPU-based) on 37 snapshots of dataset \datasetname{RTM} (sample 1 snapshot from every 100 timesteps). We find that {\ginterp} is constantly better than GPU-Lorenzo in terms of PSNR under all error bounds, gaining PSNR improvements of 2.5 to 10 dB. Moreover, attributed to the anchor point design, the PSNR from {\ginterp} even outperforms the CPU version of interpolation (implemented in SZ3~\cite{zhao2021optimizing}).

\section{Improving {\thiswork} Lossless Modules}\label{cuszinterp::design::lossless}

In this section, we detail important optimizations on {\thiswork}'s lossless encoding module, which crucially boosts the compression ratio of \ginterp-predicted {\quantcode}s.

\subsection{Tweaking Existing Huffman Coding}\label{cuszinterp::design::lossless::tweak}

Adapted from the previous practice~\cite{cusz2020}, the procedures of building a histogram of {\quantcode} for the Huffman codebook and conducting Huffman coding based on the codebook remain in the {\thiswork} pipeline.
Thanks to the more accurate prediction from {\ginterp}, the histogram contains a more centralized {\quantcode} $q$ distribution. This centralization can be modeled as follows. Recall the outlier-thresholding internal parameter $R$ mentioned in \SEC\ref{cuszinterp::background::cusz}, for each prediction with respect to the user-specified $eb$, a much smaller range, denoted by $|q| < r^\bullet < R$, corresponds to an empirical coverage of very few outliers (e.g., $<1\%$); and at the same $eb$, {\ginterp} results in a much smaller $r^\bullet$ than Lorenzo.
As reported in~\cite{tian2021revisiting}, building a Huffman tree on GPU is worthwhile only when the number of histogram entries (i.e., $r^\bullet$) is large enough. Therefore, we shift the procedure of building the Huffman codebook to CPU, with roughly an end-to-end time of 200 us. This time is \textit{excluded} in the benchmarks in \SEC\ref{cuszinterp::eval}-\{\ref{cuszinterp::eval::GBps},\enspace\ref{cuszinterp::eval::globus}\}. Admittedly, this CPU component needs to be further optimized, following approaches such as prebuilding Huffman trees~\cite{prebuilt-huffman-tree}.
On the other hand, a much smaller $r_k < r^\bullet$ further signifies this centralization and corresponds to our attempt to
allocate small thread-private buffers to cache the count of the center top-$k$ {\quantcode}s. This significantly decreases the transaction between threads' register files and the shared memory buffer. A large $k$ (or $r_k$) can incur higher register pressure; thus, for graceful degradation, $k$ can fall back to $1$, which is still helpful for highly compressible cases.

Last, though highly concentrated, {\quantcode}s in corner cases could still contain modestly deviated numbers. In those cases, we gather them as outliers and losslessly store them with trivial space and time costs using the stream compaction technique.

\subsection{Synergy of Lossless Modules: Huffman + {\bitcomp}}\label{cuszinterp::reason-why-bitcomp}

Though the Huffman-coding-only practice balances throughput and compression ratio in average cases, its compression is obstructed given that each input {\quantcode} must be represented by at least one bit; At the same time, it can still present redundancy of repeated patterns after encoding (e.g., continuous \texttt{0x00} bytes). Nevertheless, sophisticated dictionary-based encoders are either limited in throughput (e.g., GPU-LZ~\cite{zhang2023gpulz}) or compression ratio on GPU. Though previous work proposed several de-redundancy lossless encoding schemes encoding~\cite{cuszplus2021,cuszp, FZGPU} (alternative to Huffman encoding) that can surpass the one-bit-per-element limitation in extremely compressible cases, they still have unsatisfactory compression ratios in many cases, as evidenced in \FIG\ref{cuszinterp::fig::PSNR-rate}.
Therefore, we propose synergizing Huffman encoding and the subsequent repeated pattern-canceling encoding scheme, which could result in a huge gain in the compression ratio with minimal throughput overhead because this additional scheme only takes inputs of reduced sizes after Huffman encoding. %
To fit our chart of high ratio and high throughput, we selected {\bitcomp}~\cite{bitcomp} from NVIDIA, a performance-oriented encoder on GPU, after trial and error over a large variety of existing GPU-based bitstream lossless encoders,
We are aware of the proprietary nature of NVIDIA's Bitcomp; however, it only serves for demonstration purposes for effectively utilizing the created high compressibility from {\ginterp} on GPU.
In \SEC\ref{cuszinterp::eval}, we will present how {\bitcomp} helps to greatly remove the redundancy of Huffman-encoded {\quantcode}s generated by {\thiswork} data predictor with negligible computational overhead.

\section{Experimental Evaluation}\label{cuszinterp::eval}

This section presents our experimental setup and evaluation results. To systematically and convincingly evaluate {\thiswork}, experiments with diverse datasets and from various aspects of {\thiswork}, together with five other state-of-the-art error-bounded lossy compressors, are presented in this section.

\subsection{Experimental Setup}\label{cuszinterp::eval::setup}

\begin{description}[leftmargin=0.9em]
    \item[\normalfont\itshape\SQUARE{} Platforms.]
          We evaluate on
          \Circled{1} NVIDIA A100 GPUs from ALCF-ThetaGPU~\cite{testbed-thetagpu} and Purdue-Anvil~\cite{testbed-anvil} for profiling and Globus test and \Circled{2} NVIDIA A40 GPU on ANL-JLSE~\cite{testbed-jlse} for profiling. More details are in \TAB\ref{cuszinterp::tab::testbed}.

    \item[\normalfont\itshape\SQUARE{} Baselines.]
          We compare our {\thiswork} with state-of-the-art GPU-based lossy compressors, namely \cusz, cuSZp, cuSZx, cuZFP, and FZ-GPU, as baselines.

    \item[\normalfont\itshape\SQUARE{} Test datasets.]\enspace
          The experiments are based on six real-world scientific simulation datasets, detailed in \TAB\ref{cuszinterp::tab::datasets}.
          They have been reported as representative of production-level simulations in prior works~\cite{cusz2020,zhao2021optimizing,FZGPU,hpez}. Most are open data from Scientific Data Reduction Benchmarks suite~\cite{repo-sdrbench}.
\end{description}

\begin{table}[ht]
    \caption{Testbeds for our experiments.}
    \FloatBodyStyle\footnotesize
    \renewcommand{\arraystretch}{1.2}
    \begin{tabular}{ | >{\bfseries}l|c|c|c|}
    \hline
    {GPU}      & \multicolumn{2}{c|}{A100 (40GB)}       & A40 (48 GB)        \\
    \hline
    \hline
    {testbed}  & Theta-GPU    & Anvil        & JLSE         \\
    \hline
    {mem. bw.}  & \multicolumn{2}{c|}{1555 GB/s}    & 695.8  GB/s  \\
    \hline
    \begin{tabular}{@{}c@{}}
    compute (FP32) 
    \end{tabular}
    & \multicolumn{2}{c|}{19.49 TFLOPS} & 37.42 TFLOPS \\
    \hline
    CUDA version& 11.4        & 11.6        & 11.8        \\
    driver version     & 470.161.03         & 530.30.02         & 545.23.06         \\
    \hline
\end{tabular}%

    \label{cuszinterp::tab::testbed}
\end{table}

\begin{table}[ht]
    \centering
    \caption{Information of the datasets in experiments}
    \FloatBodyStyle\footnotesize
    \renewcommand{\arraystretch}{1.2}
    \begin{tabular}{|rrll|}
\hline  \multicolumn{4}{|l|}{\textbf{JHTDB}~\cite{jhtdb}: numerical simulation of turbulence.} \\
& 10 files    & dim: $512_z\times512_y\times512_x$          & total: 5 GB    \\
\hline  \multicolumn{4}{|l|}{\textbf{Miranda}~\cite{miranda}: hydrodynamics simulation.} \\
& 7 files & dim: $256_z\!\times384_y\!\times384_x$          & total: 1 GB   \\
\hline  
\multicolumn{4}{|l|}{\textbf{Nyx}~\cite{nyx}: cosmological hydrodynamics simulation.}  \\
& 6 files & dim: $512_z\!\times512_y\!\times512_x$          & total: 3.1 GB \\
\hline  
\multicolumn{4}{|l|}{\textbf{QMCPack}~\cite{qmcpack}: Monte Carlo quantum simulation.}  \\
& 1 files & dim: $(288\times115)_z\!\times69_y\!\times69_x$ & total: 612 MB \\
\hline  
\multicolumn{4}{|l|}{\textbf{RTM}~\cite{rtm}: reverse time migration for seismic imaging.} \\
& 37 files     & dim: $449_z\!\times449_y\!\times235_x$          & total: 6.5 GB \\
\hline \multicolumn{4}{|l|}{\textbf{S3D}~\cite{repo-sdrbench}: combustion process simulation.}  \\
& 11 files   & dim: $500_z\!\times500_y\!\times500_x$          & total: 5.1 GB \\
\hline
\end{tabular}

    \label{cuszinterp::tab::datasets}
\end{table}

\subsection{Evaluation Metrics}
Our evaluation is based on the following key metrics:

\newcommand{\MathCR}{\operatorname{CR}}
\newcommand{\MathSizeof}{\operatorname{sizeof}}

\begin{description}[leftmargin=0.9em]

    \item[\normalfont\itshape\SQUARE{} Fixed-error-bound compression ratio (CR).]
          CR is the original input size divided by the compressed size. %
          
    \item[\normalfont\itshape\SQUARE{} Rate-distortion graph.]
          This graph shows the compression bit rate and the decompression data PSNR for compressors. The bit rate $b$ is the average of bits in the compressed data for each input element (i.e., $32\times$ the reciprocal of CR).
          
    \item[\normalfont\itshape\SQUARE{} Fixed-CR visualization.]
          The visual qualities of the reconstructed data from all the compressors at the same CR.

    \item[\normalfont\itshape\SQUARE{} Throughput.]
          Compression and decompression throughputs of all the compressors in GB/s.

    \item[\normalfont\itshape\SQUARE{} Data transfer time.]
          We perform distributed and parallel data transfer tests in the form of lossy compression archives (from all compressors) on multiple supercomputers.
\end{description}

\subsection{Evaluation Results and Analysis}

\subsubsection{Compression Ratios.}\label{cuszinterp::eval::CR}

We compress the datasets under fixed error bounds that are commonly used and list all the results in \TAB\ref{cuszinterp::tab::CR-bitcomp} (no results for cuZFP as it does not support absolute error-bounding, and no results for cuSZx on dataset \datasetname{Nyx} for its runtime errors). \TAB\ref{cuszinterp::tab::CR-bitcomp} presents the two-fold achievement of our novel pipeline.
\Circled{1} In the left half, we present the compression ratios \textit{without} an extra lossless module; even so, {\thiswork} outperforms others in 14 of 18 cases, exhibiting 10\% to 30\% advantages over the second-bests (column 6).
\Circled{2} In the right half, we examine the full pipeline integrating {\bitcomp}. For fairness, we apply {\bitcomp} to all other compressors' outputs.
\emph{First}, the full pipeline sees a significant gain in compression ratio from the one without {\bitcomp} (i.e., comparing column v with 5). \emph{Second}, {\thiswork} achieves the highest compression ratios in all 18 cases. {\thiswork} expands advantage over the second-best with {\bitcomp} enabled (column vi) and tops at 476\%. These note that {\ginterp} creates more profound compressibility and is more attuned to the additional pass of lossless encoding than any other compressors.
\textit{Last}, it is worth noticing that, for all results in \TAB\ref{cuszinterp::tab::CR-bitcomp}, the decompression PSNR of {\thiswork} is also much higher than other compressors. If we compare the compression ratio under the same PSNR instead of the same error bound, the improvements from {\thiswork} will be amplified (to be detailed in \SEC\ref{cuszinterp::eval::rate-distortion}).

\begin{table}[ht]
    \caption {Compression ratios without (columns 1 to 6) and with {\bitcomp} (cols. i to vi) at error bounds \texttt{1e-2}, \texttt{1e-3}, and \texttt{1e-4}. The best CRs are boldface, and the second-best are underlined.}
    \FloatBodyStyle
    \newcommand{\TabDataName}[1]{\noindent\rotatebox{90}{\bfseries\scriptsize\scalebox{0.9}{\noindent#1}}}
    \setlength{\tabcolsep}{2pt}
    \scriptsize\color{gray}
    \renewcommand{\arraystretch}{1.3}
    \FBOXDEBUG{\begin{adjustbox}{width=\columnwidth}
            \begin{tabular}{ >{\color{black}}l@{} >{\color{black}}c| *{2}{c |rrrrr| >{\color{black}}r|} }
  \multicolumn{1}{l}{\rotatebox{90}{\textbf{Dataset}}} &
  \multicolumn{1}{r}{\rotatebox{90}{\textbf{epsilon}}} &
  \multicolumn{1}{r}{} &
  \multicolumn{1}{r}{\rotatebox{90}{\textbf{\cusz}}}    &
  \multicolumn{1}{r}{\rotatebox{90}{\textbf{cuSZp}}}   &
  \multicolumn{1}{r}{\rotatebox{90}{\textbf{cuSZx}}}   &
  \multicolumn{1}{r}{\rotatebox{90}{\textbf{FZ-GPU}}}  &
  \multicolumn{1}{r}{\rotatebox{90}{\color{black}\textbf{\thiswork}}}  &
  \multicolumn{1}{r}{\rotatebox{90}{\color{black}\textbf{Advant.\%}}} &
  \multicolumn{1}{r}{} &
  \multicolumn{1}{r}{\rotatebox{90}{\textbf{\cusz}}}    &
  \multicolumn{1}{r}{\rotatebox{90}{\textbf{cuSZp}}}   &
  \multicolumn{1}{r}{\rotatebox{90}{\textbf{cuSZx}}}   &
  \multicolumn{1}{r}{\rotatebox{90}{\textbf{FZ-GPU}}}  &
  \multicolumn{1}{r}{\rotatebox{90}{\color{black}\textbf{\thiswork}}}  &
  \multicolumn{1}{r}{\rotatebox{90}{\color{black}\textbf{Advant.\%}}}
  \\
  \cline{1-2}\cline{4-9}\cline{11-16}
  \multirow{3}{*}{\TabDataName{JHTDB}}                 & 1e--2 && \SecondbestCr{26.6} & 10.3                & 3.0               & 12.1              & \NoBcBestCr{29.3}   & 10.2   &  & \SecondbestCr{27.8} & 19.9                 & 3.1  & 18.0 & \BestCr{132.0} & 374.8 \\
                                                       & 1e--3 && \SecondbestCr{17.7} & 5.4                 & 2.5               & 9.9               & \NoBcBestCr{25.2}   & 42.4   &  & \SecondbestCr{17.7} & 6.0                  & 2.5  & 11.5 & \BestCr{34.8}  & 96.6  \\
                                                       & 1e--4 && \SecondbestCr{10.7} & 3.5                 & 1.8               & 6.4               & \NoBcBestCr{13.3}   & 24.3   &  & \SecondbestCr{10.7} & 3.6                  & 1.8  & 7.8  & \BestCr{13.3}  & 24.3  \\

   \cline{1-2}\cline{4-9}\cline{11-16}
  \multirow{3}{*}{\TabDataName{Miranda}}               & 1e--2 && 27.1                & 16.8                & 7.9               & \NoBcBestCr{30.6} & \SecondbestCr{28.5} & --6.9  &  & \SecondbestCr{67.4} & 18.7                 & 8.1  & 43.9 & \BestCr{174.0} & 158.2 \\
                                                       & 1e--3 && \SecondbestCr{22.9} & 9.6                 & 5.1               & 19.2              & \NoBcBestCr{26.3}   & 14.8   &  & \SecondbestCr{38.5} & 10.7                 & 5.2  & 27.1 & \BestCr{77.2}  & 100.5 \\
                                                       & 1e--4 && \SecondbestCr{15.3} & 6.0                 & 3.6               & 11.8              & \NoBcBestCr{19.5}   & 27.5   &  & \SecondbestCr{19.8} & 6.7                  & 3.7  & 15.4 & \BestCr{34.3}  & 73.2  \\
  \cline{1-2}\cline{4-9}\cline{11-16}
  \multirow{3}{*}{\TabDataName{Nyx}}                   & 1e--2 && \NoBcBestCr{30.2}   & 20.3                & N/A               & 25.3              & \SecondbestCr{29.6} & --2.0  &  & 71.6                & \SecondbestCr{95.9}  & N/A  & 84.5 & \BestCr{256.0} & 166.9 \\
                                                       & 1e--3 && \SecondbestCr{23.9} & 9.6                 & N/A               & 14.4              & \NoBcBestCr{28.0}   & 17.2   &  & \SecondbestCr{34.4} & 19.0                 & N/A  & 26.2 & \BestCr{66.1}  & 92.2  \\
                                                       & 1e--4 && \SecondbestCr{15.3} & 5.7                 & N/A               & 8.4               & \NoBcBestCr{18.7}   & 22.2   &  & \SecondbestCr{17.9} & 7.5                  & N/A  & 12.3 & \BestCr{25.1}  & 40.2  \\
 
  \cline{1-2}\cline{4-9}\cline{11-16}
  \multirow{3}{*}{\TabDataName{Qmcpack}}               & 1e--2 && \SecondbestCr{28.6} & 22.2                & 3.3               & 19.0              & \NoBcBestCr{29.3}   & 2.4    &  & \SecondbestCr{46.0} & 38.7                 & 3.3  & 30.3 & \BestCr{168.0} & 265.2 \\
                                                       & 1e--3 && \SecondbestCr{20.9} & 10.1                & 2.5               & 12.1              & \NoBcBestCr{27.7}   & 32.5   &  & \SecondbestCr{23.7} & 11.5                 & 2.5  & 14.7 & \BestCr{78.7}  & 232.1 \\
                                                       & 1e--4 && \SecondbestCr{14.8} & 5.6                 & 1.9               & 8.3               & \NoBcBestCr{22.6}   & 52.7   &  & \SecondbestCr{15.3} & 5.8                  & 1.9  & 10.2 & \BestCr{34.6}  & 126.1 \\
\cline{1-2}\cline{4-9}\cline{11-16}
  \multirow{3}{*}{\TabDataName{RTM}}                   & 1e--2 && 28.7                & \SecondbestCr{41.6} & \NoBcBestCr{53.7} & 32.0              & 28.8                & --46.4 &  & 84.1                & \SecondbestCr{100.5} & 70.4 & 69.7 & \BestCr{234.0} & 132.8 \\
                                                       & 1e--3 && 24.7                & 19.8                & \NoBcBestCr{30.7} & 20.9              & \SecondbestCr{27.4} & --10.7 &  & \SecondbestCr{50.2} & 31.1                 & 38.8 & 35.5 & \BestCr{96.2}  & 91.6  \\
                                                       & 1e--4 && \SecondbestCr{17.7} & 10.7                & 17.4              & 12.1              & \NoBcBestCr{21.5}   & 21.5   &  & \SecondbestCr{26.6} & 14.0                 & 21.4 & 18.4 & \BestCr{45.4}  & 70.7  \\
  \cline{1-2}\cline{4-9}\cline{11-16}
  
  \multirow{3}{*}{\TabDataName{S3D}}                   & 1e--2 && \SecondbestCr{28.0} & 7.2                 & 19.5              & 15.5              & \NoBcBestCr{29.5}   & 5.4    &  & \SecondbestCr{42.5} & 18.9                 & 19.9 & 25.6 & \BestCr{245.0} & 476.5 \\
                                                       & 1e--3 && \SecondbestCr{23.3} & 4.5                 & 9.3               & 11.8              & \NoBcBestCr{28.8}   & 23.6   &  & \SecondbestCr{28.3} & 8.8                  & 9.5  & 16.1 & \BestCr{137.0} & 384.1 \\
                                                       & 1e--4 && \SecondbestCr{17.3} & 3.1                 & 5.0               & 9.0               & \NoBcBestCr{26.0}   & 50.3   &  & \SecondbestCr{19.0} & 5.0                  & 5.1  & 11.6 & \BestCr{58.2}  & 206.3 \\
  
  \cline{1-2}\cline{4-9}\cline{11-16}
  
\multicolumn{3}{c|}{} & \multicolumn{6}{c|}{\color{black}\bfseries without {\bitcomp}} & \multicolumn{1}{c|}{} & \multicolumn{6}{c|}{\color{black}\bfseries {with {\bitcomp}} } \\
\multicolumn{2}{c}{col. no.} & \multicolumn{1}{c}{} &
1 & 2 & 3 & 4       & \multicolumn{1}{r}{5} & \multicolumn{1}{r}{6} & 
\multicolumn{1}{c}{} & 
i & ii & iii & iv   & \multicolumn{1}{r}{v} & \multicolumn{1}{r}{vi} \\
\end{tabular}

        \end{adjustbox}}

    \label{cuszinterp::tab::CR-bitcomp}
\end{table}

\subsubsection{Compression Rate-Distortion.}\label{cuszinterp::eval::rate-distortion}

\begin{figure*}
    \begin{subfigure}{\linewidth}
        \tikz[inner sep=0pt]{
            \node{\includegraphics[width=\linewidth, trim={0 8mm 0 1mm}, clip]{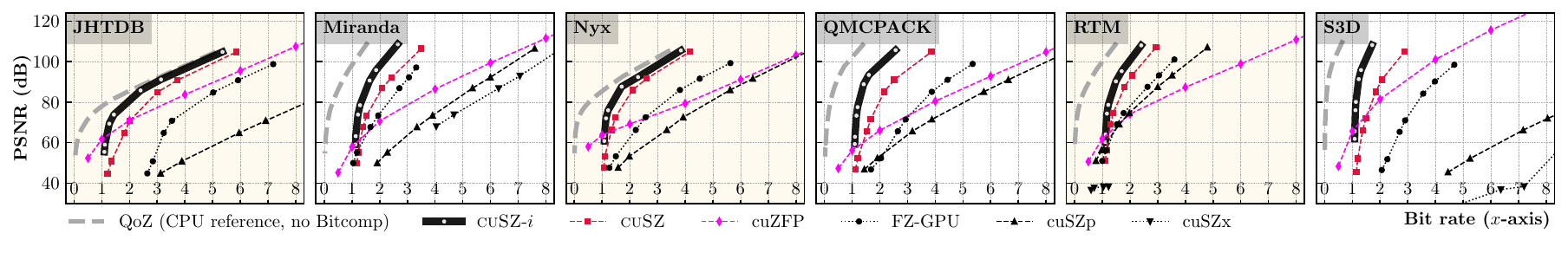}};
        }
        \tikz[inner sep=0pt]{
            \node{\includegraphics[width=\linewidth, trim={0 3mm 0 1mm}, clip]{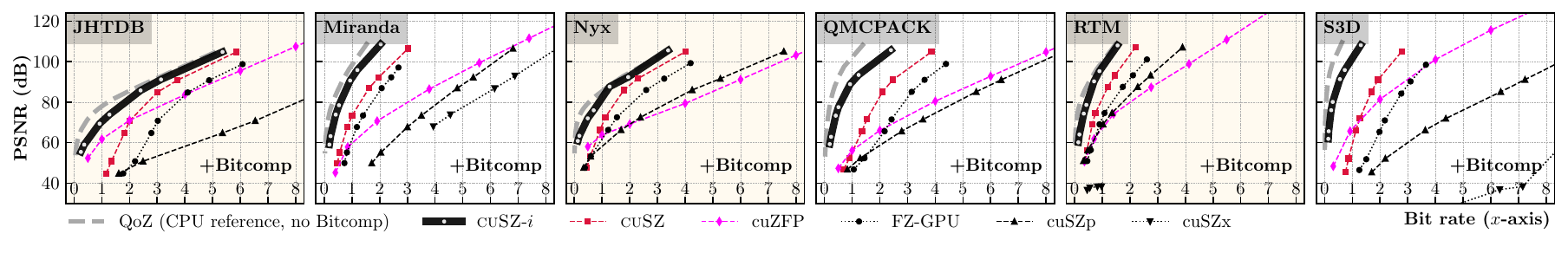}};
        }
        \vspace{-6mm}
        \caption{(Bit rate)-PSNR graphs from compression results without (top row) and with {\bitcomp} (bottom row) encoding pass.}
        \label{cuszinterp::fig::PSNR-rate}
    \end{subfigure}

    \begin{subfigure}{\linewidth}
        \vspace{1mm}
        \fontfamily{cmr}\selectfont
        \tikz{
            \node[inner sep=0pt](a){\includegraphics[width=\linewidth, trim={0 3mm 0 2mm}, clip]{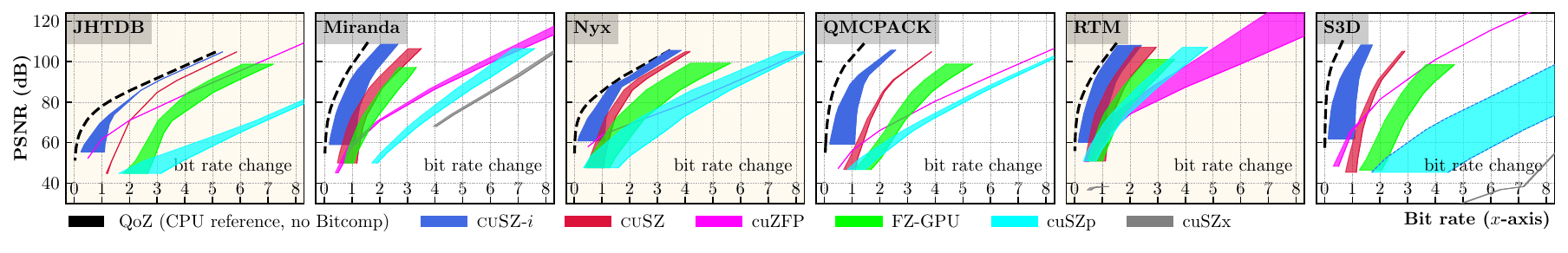}};
            \draw[-latex, densely dotted, thin, color=black] ([xshift=-.16in]a.east) -- node[above, align=right, scale=.45, yshift=.2ex, xshift=1em]{fixed PSNR,\\ smaller bit rate} ([xshift=-.6in]a.east);
        }%
        \vspace{-2mm}
        \caption{(Bit rate)-PSNR \textit{leftward change} due to the fixed PSNR and the smaller bit rate from the extra pass of lossless encoding (\bitcomp).}
        \label{cuszinterp::fig::rate-transition}
    \end{subfigure}

    \label{cuszinterp::fig::PSNR-rate-grand}
    \caption{Compression (bit rate)-PSNR graphs on six datasets. A curve toward \textit{upperleft} indicates \emph{both} advantageous quality and compression ratios.}
    \vspace{-2ex}
\end{figure*}

Both Interpolation-based data predictor and {\bitcomp} contribute to better compression rate-distortion trait.
To exhibit the separate contributions, we present the rate-PSNR curves on 6 datasets in 2 parallel series in \FIG\ref{cuszinterp::fig::PSNR-rate}: without and with {\bitcomp}.
When no extra lossless module is appended, {\thiswork} has already achieved the best rate-distortion, attributed to the effective {\ginterp} data predictor. On dataset \datasetname{JHTDB} under a PSNR of 70~dB, or on dataset \datasetname{QMCPACK} under a PSNR of 80~dB, {\thiswork} shows 60\% to 80\% advantages in compression ratio over the second-best cuZFP/\cusz. In the with-Bitcomp results, {\thiswork} has far better compression ratios than any other, achieving roughly 100\% to 500\% advantages under the same PSNRs in low-bit-rate cases. In high-bit-rate cases, it also promotes considerable reduction rates. The great extent of compression boost by {\thiswork} originates from its high adaptability to {\bitcomp} since its {\ginterp} results in highly compressible error quant-codes, which is visualized as the fixed-PSNR bit rate change in auxiliary \FIG\ref{cuszinterp::fig::rate-transition}. As a reference baseline in \FIG\ref{cuszinterp::fig::PSNR-rate}, we also include the rate-distortion of QoZ~\cite{qoz},
which reflects the latest interpolation-based art on the CPU platform and one design basis of {\thiswork}.
We conclude that 1) CPU-based QoZ still features a better compression ratio than {\thiswork} due to larger interpolation blocks and more effective lossless modules, but {\thiswork} features far better throughputs; 2) nevertheless, {\thiswork} (with \bitcomp) has, for the first time, approximated the best-in-class CPU interpolation-based compressor (i.e., QoZ) in rate-distortion trait.

\subsubsection{Case Study of Decompression Visualization.}\label{cuszinterp::eval::visual}

\begin{figure*}
    \vspace{10mm}
    \FBOXDEBUG{\includegraphics[trim={2mm 0mm 2mm 9mm}, width=\linewidth]{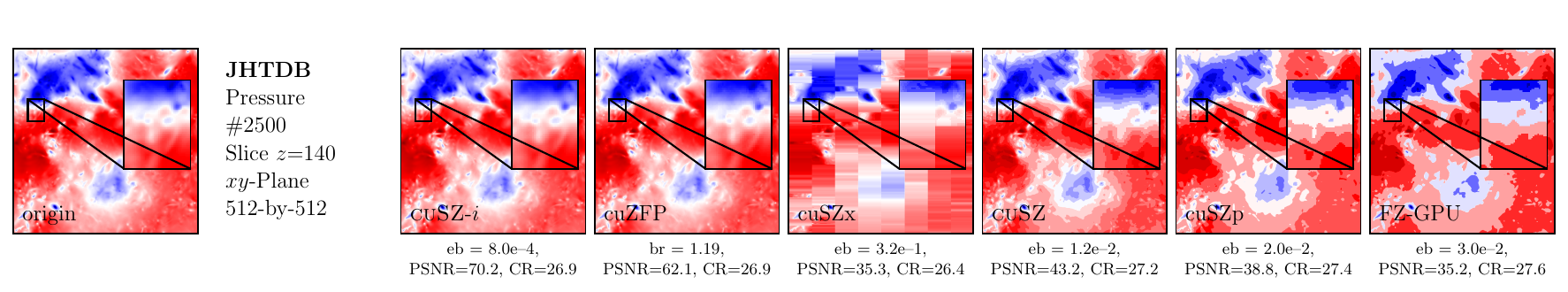}}
    \FBOXDEBUG{\includegraphics[trim={2mm 0mm 2mm 9mm}, width=\linewidth]{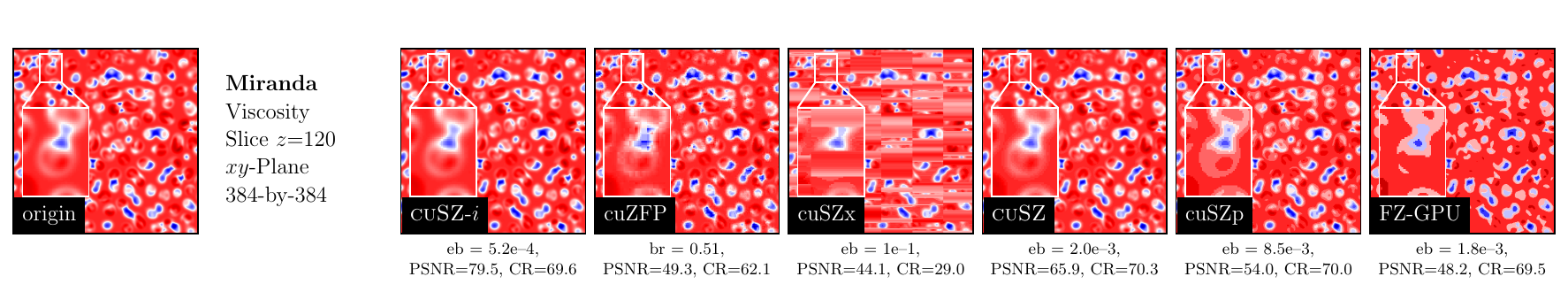}}
    \FBOXDEBUG{\includegraphics[trim={2mm 0mm 2mm 9mm}, width=\linewidth]{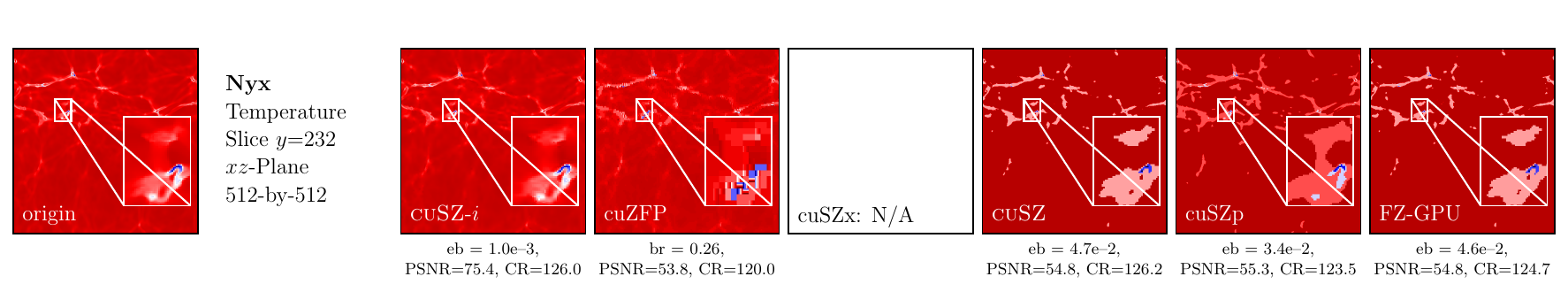}}
    \FBOXDEBUG{\includegraphics[trim={2mm 0mm 2mm 9mm}, width=\linewidth]{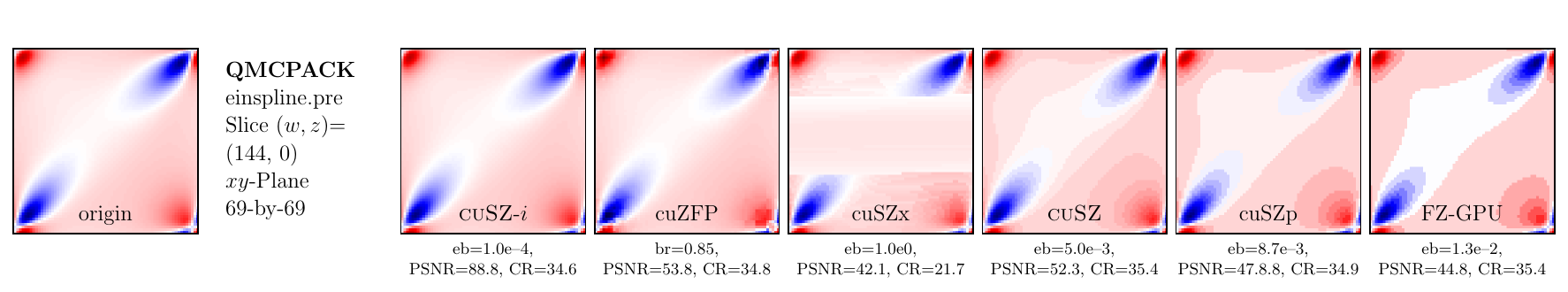}}
    \FBOXDEBUG{\includegraphics[trim={2mm 0mm 2mm 2mm}, width=\linewidth]{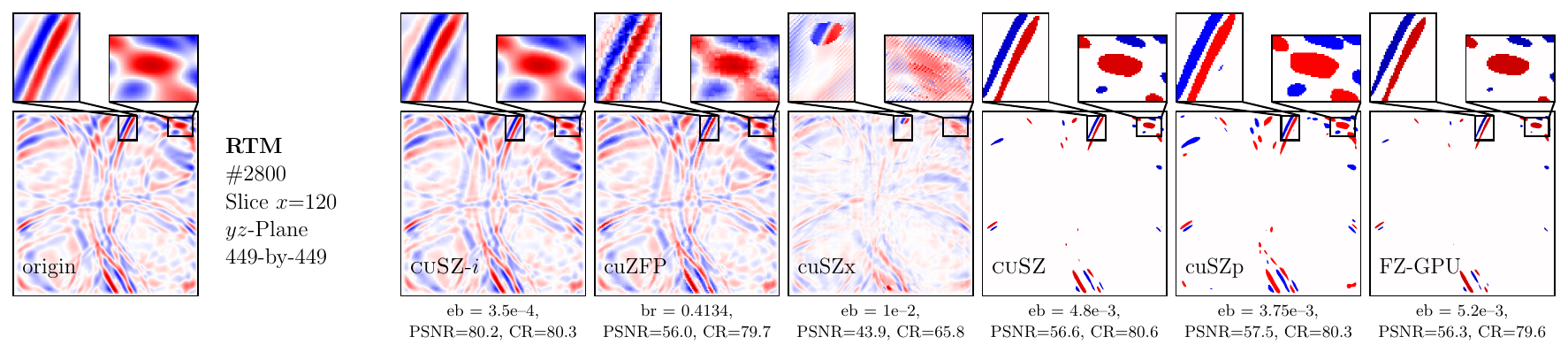}}
    \FBOXDEBUG{\includegraphics[trim={2mm 0mm 2mm 9mm}, width=\linewidth]{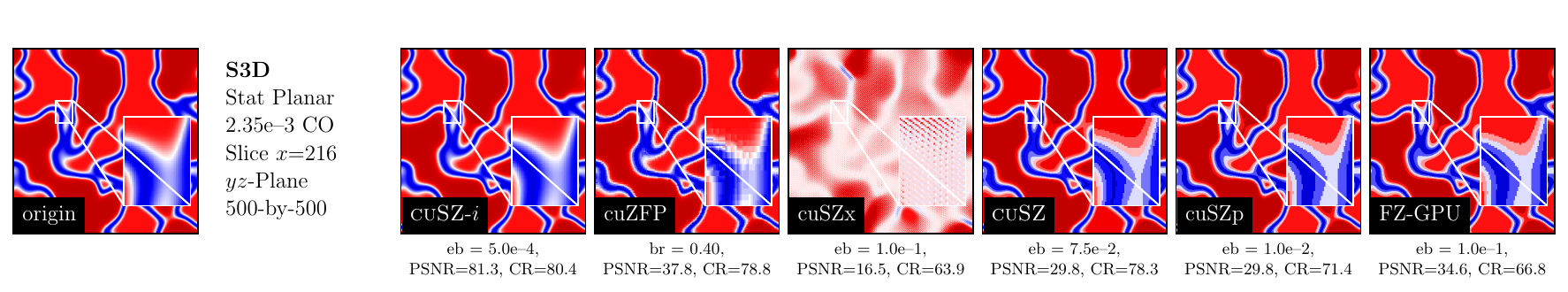}}

    \caption{Showcases of visualized decompressed data, with key subregions zoomed in. For each snapshot, we align the (with-Bitcomp) compression ratio for every compressor with a fixed values. In addition, {\cusz}, FZ-GPU, and cuSZp share the same Lorenzo predictor, exhibiting simiar data visualizations. Hence, they are grouped as the last three.}
    \label{cuszinterp::fig::visual-all}
\end{figure*}

To further verify the high compression quality of {\thiswork}, we visualize certain data decompression snapshots from the datasets.
\FIG\ref{cuszinterp::fig::visual-all} shows the original and decompression visualization of two data snapshots by six compressors, together with a note of the compression ratios, error bounds, and PSNRs. For each set of visualizations on the same snapshot, we align the compression ratios to the same value (e.g., $\sim$ 27 for \datasetname{JHTDB}).

Under a fixed compression ratio, {\thiswork} has the best and closest-to-the-original visualization quality from the decompressed data in all cases. In contrast, all other compressors have exhibited severe visualization artifacts. On the visualized decompressions of \datasetname{JHTDB}, {\thiswork} achieved a PSNR of 70.2~dB, which is 8~dB better than the second-best cuZFP. On data snapshot \datafieldname{CO} of \datasetname{S3D}, under the same compression ratio, around 80~dB. The decompression PSNR of {\thiswork} gets a value of 81.3~dB, notably outperforming all other existing compressors (the second-best is only 37.8~dB).

\subsubsection{Compression Throughputs.}\label{cuszinterp::eval::GBps}

\begin{figure*}[ht]
    \begin{subfigure}{\linewidth}
        \centering
        \FBOXDEBUG{\includegraphics[width=\linewidth, trim={0 8pt 0 3pt}, clip]{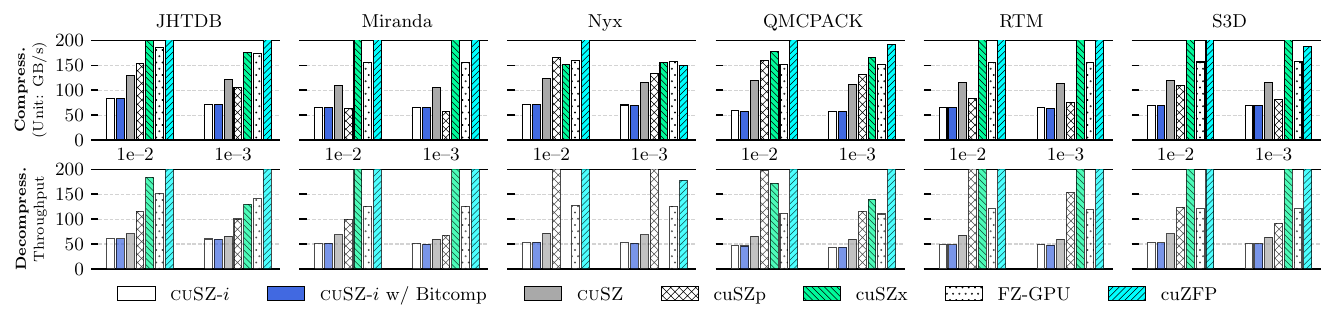}}
        \vspace{-6.0mm}
        \caption{Compression and decompression throughputs on NVIDIA A100.}
        \label{cuszinterp::fig::A100-GBps}
    \end{subfigure}%
    \vspace{1mm}
    \begin{subfigure}{\linewidth}
        \centering
        \FBOXDEBUG{\includegraphics[width=\linewidth, trim={0 8pt 0 3pt}, clip]{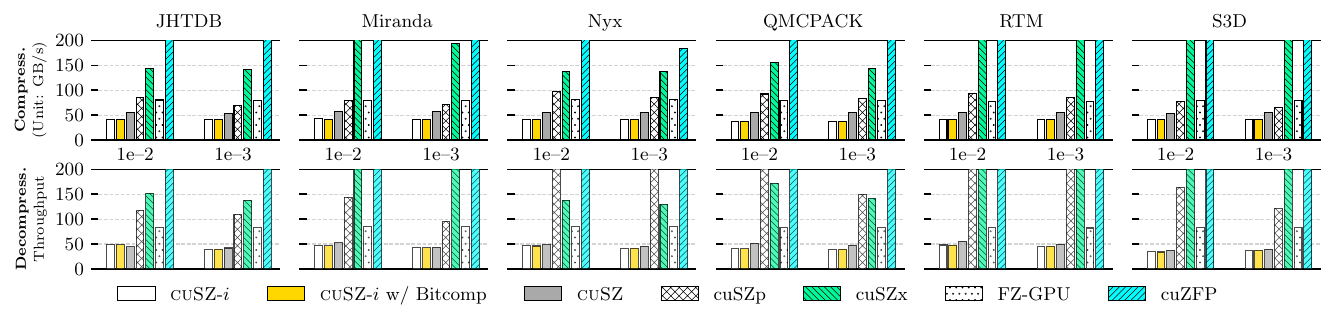}}
        \vspace{-6.0mm}
        \caption{Compression and decompression throughputs on NVIDIA A40.}
        \label{cuszinterp::fig::A40-GBps}
    \end{subfigure}

    \caption{Compression and decompression throughputs on NVIDIA A100 ({top} row) and A40 ({bottom} row) for {\thiswork}, {\thiswork} with {\bitcomp}, {\cusz}, cuZFP, cuSZp, cuSZx, and FZ-GPU. cuZFP's throughput corresponds to a similar average PSNR with {\thiswork}.}
    \label{cuszinterp::fig::A100-and-A40-GBps}
    \vspace{-2.0mm}
\end{figure*}

After discussing the compression ratio and quality of the compressors, we would like to address the concerns regarding the throughputs.
We profiled the compression throughputs of the GPU-based lossy compressors by measuring the kernel execution time with NVIDIA Nsight System on two NVIDIA GPUs: A100 on ThetaGPU and A40 on JLSE. \FIG\ref{cuszinterp::fig::A100-GBps} presents the compression and decompression throughputs of all 6 GPU-based compressors under two different error bounds, \texttt{1e-2} and \texttt{1e-3}, on the A100 on ThetaGPU. Specifically, \underline{\textsf{\thiswork}} indicates the {\thiswork} pipeline without {\bitcomp}, and \underline{\textsf{{\thiswork} w/ Bitcomp}} indicates the proposed full pipeline of higher compression ratios. We observe that adding {\bitcomp} brings negligible overhead to compression throughputs. Also, {\thiswork}'s throughput is at the same magnitude as {\cusz} and cuSZp. Specifically, it has approximately 60\% of {\cusz}'s compression throughput and 80\% to 90\% of {\cusz}'s decompression throughput. Moreover, \FIG\ref{cuszinterp::fig::A40-GBps} presents profiling results on A40 on JLSE. Here, {\thiswork} performs closer to {\cusz}, reaching 70\% to 80\% of {\cusz}'s compression throughput, and nearly the same as {\cusz}'s decompression throughput.

Due to the intrinsic data dependency and more sophisticated computational operations in spline interpolation kernels, interpolation-based {\thiswork} is inevitably slower than Lorenzo-based {\cusz} and its derivatives. However, the kernel throughput of {\thiswork} is still on the same magnitude as {\cusz}, and it can be further improved to better serve as a component for real-time processing. Moreover, although cuSZx, cuZFP, and FZ-GPU have relatively higher throughput than {\thiswork}, they present much lower compression ratios than {\thiswork}'s, inadequate to many use cases in which data storage and transfer costs need to be minimized. Next, we will see how {\thiswork} can be advantageous in real-use tasks that are even sensitive to compression throughputs, and when considering the achieved data decompression quality and the compression ratio simultaneously, {\thiswork} have established the Pareto front in scenarios of transferring data over the bandwidth-limited channels.

\begin{figure}[ht]
    \centering
    \FBOXDEBUG{\includegraphics[width=\linewidth]{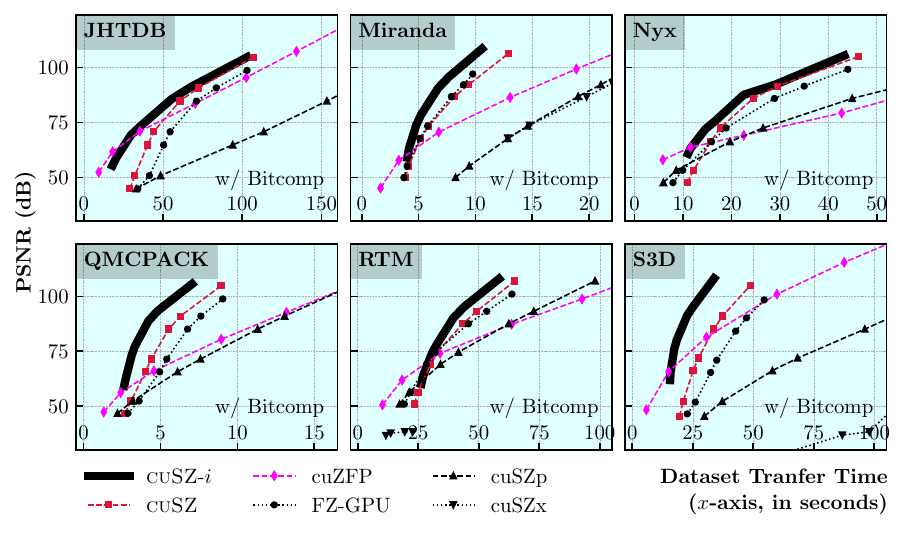}}
    \vspace*{-5mm}
    \caption{(Transfer time)-PSNR graphs of compressors on six datasets. A curve toward \textit{upperleft} indicates the advantageous performance in \textit{both} better quality and shorter transfer time.}
    \label{cuszinterp::fig::PSNR-transfer-time}
\end{figure}

\subsubsection{Case Study: Distributed Lossy Data Transmission.}\label{cuszinterp::eval::globus}

In this part, we conduct a practical case study of distributed lossy data transmission as discussed in \cite{globus-compression}. Initially, a distributed scientific database is deployed across several supercomputers. Rapid data transfer and access between two distributed machines within this database system are desired. Rather than transfer the original exascale data due to an impractical amount of time, the database system can transfer the significantly smaller compressed data. To this end, an error-bounded lossy compressor operates on the source and destination machines. The overall time required for this process is the sum of local data I/O time, the data compression and decompression time, and the time for transferring the compressed data between the machines. Since the data transfer between remote machines will be performed under relatively low bandwidths, a data compressor with both high ratios and high performance (therefore, the GPU-based is preferred) is required for an optimized data transfer efficiency.

\FIG\ref{cuszinterp::fig::PSNR-transfer-time} shows the data-transfer cost in time and decompression data PSNR for this task. Specifically, the source and destination machines are ALCF Theta-GPU and Purdue RCAC Anvil, respectively. Since the local I/O time is irrelevant to this scenario, we compute the overall time only from data compression/decompression time and the data-transfer time of distributed compressed data. Managed by Globus~\cite{globus2}, the data-transfer bandwidth between the two machines is approximately 1 GB/s. With the full pipeline (including {\bitcomp}) fairly applied to all compressors to further reduce data-transfer costs, {\cuszinterp} has the best-in-class time cost on all high-quality data-transfer cases (PSNR $\ge$ 70 dB), as indicated in \FIG\ref{cuszinterp::fig::PSNR-transfer-time}. {\cuszinterp} is also very competitive on specific data transfer tasks with low-quality data. For dataset \datasetname{QMCPACK}, {\cuszinterp} can reduce around 30\% of the time cost of the second-best {\cusz} at a PSNR of 90 dB. For dataset \datasetname{S3D}, it can reduce around 40\% of the time cost of the second-best cuZFP at a {PSNR} of 80 dB. Overall, it is evident that the high compression ratio and quality of {\cuszinterp} have gracefully compensated for its speed degradation, implying significance in a great variety of real-world use cases.

\section{Conclusion and Future Work}
\label{cuszinterp::conclusion}
In this work, we propose {\thiswork}, an efficient GPU-based scientific error-bounded lossy compressor, which exhibits the best compression ratio and quality compared with other related works. With a highly parallelized interpolation-based data predictor and a lightweight auto-tuning mechanism, {\thiswork} achieves state-of-the-art compression quality under the same error bounds or compression ratios rooted in its high data prediction accuracy.
With an additional lossless module (NVIDIA Bitcomp),
{\thiswork} further outperforms other GPU-based scientific lossy compressors in compression ratios by a significant advantage, approaching CPU-based compressors with far higher throughputs than CPU-based ones. In the case studies, {\thiswork} presents excellent decompression data visualization that prevails over existing state-of-the-art GPU-based lossy compressors and reduces the distributed lossy data transmission time to a considerable extent for distributed scientific databases.

    {\thiswork} has a few limitations, e.g., its compression speeds are slower than other GPU-based compressors, its interpolation-based prediction still has lower accuracy than the CPU-based interpolators, and its lossless module involves CPU and is partially dependent on the NVIDIA GPU ecosystem. In the future, we will endeavor to address those limitations, working on improving its speeds, prediction accuracy, and compatibility with more GPU architectures such as AMD and Intel GPUs.

\section*{Acknowledgments}
This research was supported by the U.S. Department of Energy, Office of Science, Advanced Scientific Computing Research (ASCR), under contracts \texttt{DE-AC02-06CH11357}. This work was also supported by the National Science Foundation (Grant Nos. \texttt{2003709}, \texttt{2303064}, \texttt{2104023}, \texttt{2247080}, \texttt{2247060}, \texttt{2312673}, \texttt{2311875}, and \texttt{2311876}).
We also acknowledge the computing resources provided by Argonne Leadership Computing Facility (ALCF) and Advanced Cyberinfrastructure Coordination Ecosystem---Purdue Anvil through Services \& Support (ACCESS).

\newpage
\renewcommand*{\bibfont}{\footnotesize}
\printbibliography[]

@inproceedings{tac2022,
  author    = {Wang, Daoce and Pulido, Jesus and Grosset, Pascal and Jin, Sian and Tian, Jiannan and Ahrens, James and Tao, Dingwen},
  title     = {{TAC}: Optimizing Error-Bounded Lossy Compression for Three-Dimensional Adaptive Mesh Refinement Simulations},
  year      = {2022},
  isbn      = {9781450391993},
  publisher = {Association for Computing Machinery},
  address   = {New York, NY, USA},
  url       = {https://doi.org/10.1145/3502181.3531458},
  doi       = {10.1145/3502181.3531458},
  booktitle = {Proceedings of the 31st International Symposium on High-Performance Parallel and Distributed Computing},
  pages     = {135–147},
  numpages  = {13},
  location  = {Minneapolis, MN, USA},
  series    = {HPDC '22}
}

@inproceedings{amric2023,
  author    = {Wang, Daoce and Pulido, Jesus and Grosset, Pascal and Tian, Jiannan and Jin, Sian and Tang, Houjun and Sexton, Jean and Di, Sheng and Zhao, Kai and Fang, Bo and Luki\'{c}, Zarija and Cappello, Franck and Ahrens, James and Tao, Dingwen},
  title     = {{AMRIC}: A Novel In Situ Lossy Compression Framework for Efficient I/O in Adaptive Mesh Refinement Applications},
  year      = {2023},
  isbn      = {9798400701092},
  publisher = {Association for Computing Machinery},
  address   = {New York, NY, USA},
  url       = {https://doi.org/10.1145/3581784.3613212},
  doi       = {10.1145/3581784.3613212},
  booktitle = {Proceedings of the International Conference for High Performance Computing, Networking, Storage and Analysis},
  articleno = {44},
  numpages  = {15},
  location  = {Denver, CO, USA},
  series    = {SC '23}
}

@article{tacplus2024,
  author   = {Wang, Daoce and Pulido, Jesus and Grosset, Pascal and Jin, Sian and Tian, Jiannan and Zhao, Kai and Ahrens, James and Tao, Dingwen},
  journal  = {IEEE Transactions on Parallel and Distributed Systems},
  title    = {{TAC+}: Optimizing Error-Bounded Lossy Compression for 3D AMR Simulations},
  year     = {2024},
  volume   = {35},
  number   = {3},
  pages    = {421-438},
  doi      = {10.1109/TPDS.2023.3339474}
}

@inproceedings{prebuilt-huffman-tree,
  author    = {Shah, Milan and Yu, Xiaodong and Di, Sheng and Becchi, Michela and Cappello, Franck},
  title     = {Lightweight Huffman Coding for Efficient GPU Compression},
  year      = {2023},
  isbn      = {9798400700569},
  publisher = {Association for Computing Machinery},
  doi       = {10.1145/3577193.3593736},
  booktitle = {Proceedings of the 37th International Conference on Supercomputing},
  pages     = {99–110},
  numpages  = {12},
  location  = {Orlando, FL, USA},
  series    = {ICS '23}
}

@inproceedings{hybrid-sz,
  author    = {Liang, Xin and Di, Sheng and Li, Sihuan and Tao, Dingwen and Nicolae, Bogdan and Chen, Zizhong and Cappello, Franck},
  title     = {Significantly improving lossy compression quality based on an optimized hybrid prediction model},
  year      = {2019},
  isbn      = {9781450362290},
  publisher = {Association for Computing Machinery},
  address   = {New York, NY, USA},
  url       = {https://doi.org/10.1145/3295500.3356193},
  doi       = {10.1145/3295500.3356193},
  booktitle = {Proceedings of the International Conference for High Performance Computing, Networking, Storage and Analysis},
  articleno = {33},
  numpages  = {26},
  location  = {Denver, Colorado},
  series    = {SC '19}
}

@inproceedings{shengdi-quantum-sim-2019,
  author    = {Wu, Xin-Chuan and Di, Sheng and Dasgupta, Emma Maitreyee and Cappello, Franck and Finkel, Hal and Alexeev, Yuri and Chong, Frederic T.},
  title     = {Full-state quantum circuit simulation by using data compression},
  year      = {2019},
  isbn      = {9781450362290},
  publisher = {Association for Computing Machinery},
  doi       = {10.1145/3295500.3356155},
  booktitle = {Proceedings of the International Conference for High Performance Computing, Networking, Storage and Analysis},
  articleno = {80},
  numpages  = {24},
  location  = {Denver, Colorado},
  series    = {SC '19}
}

@misc{testbed-anvil,
  howpublished = {\url{https://www.rcac.purdue.edu/anvil}},
  title        = {{Anvil - Purdue RCAC}}
}

@misc{cuZFP,
  author       = {{cuZFP}},
  howpublished = {\url{https://github.com/LLNL/zfp/tree/develop/src/cuda_zfp}},
  year         = {2019},
  note         = {Online}
}

@article{son2014data,
  title   = {Data compression for the exascale computing era-survey},
  author  = {Son, Seung Woo and Chen, Zhengzhang and Hendrix, William and Agrawal, Ankit and Liao, Wei-keng and Choudhary, Alok},
  journal = {Supercomputing Frontiers and Innovations},
  volume  = {1},
  number  = {2},
  pages   = {76--88},
  year    = {2014}
}

@article{use-case-Franck,
  title     = {Use cases of lossy compression for floating-point data in scientific data sets},
  author    = {Cappello, Franck and Di, Sheng and Li, Sihuan and Liang, Xin and Gok, Ali Murat and Tao, Dingwen and Yoon, Chun Hong and Wu, Xin-Chuan and Alexeev, Yuri and Chong, Frederic T},
  journal   = {The International Journal of High Performance Computing Applications},
  volume    = {33},
  number    = {6},
  pages     = {1201--1220},
  year      = {2019},
  publisher = {SAGE Publications Sage UK: London, England}
}

@article{zfp,
  title     = {Fixed-rate compressed floating-point arrays},
  author    = {Lindstrom, Peter},
  journal   = {IEEE Transactions on Visualization and Computer Graphics},
  volume    = {20},
  number    = {12},
  pages     = {2674--2683},
  year      = {2014},
  publisher = {IEEE}
}

@inproceedings{SPERR,
  title        = {Lossy scientific data compression with SPERR},
  author       = {Li, Shaomeng and Lindstrom, Peter and Clyne, John},
  booktitle    = {2023 IEEE International Parallel and Distributed Processing Symposium (IPDPS)},
  pages        = {1007--1017},
  year         = {2023},
  organization = {IEEE}
}

@article{ballester2019tthresh,
  title     = {{TTHRESH}: Tensor compression for multidimensional visual data},
  author    = {Ballester-Ripoll, Rafael and Lindstrom, Peter and Pajarola, Renato},
  journal   = {IEEE transactions on visualization and computer graphics},
  volume    = {26},
  number    = {9},
  pages     = {2891--2903},
  year      = {2019},
  publisher = {IEEE}
}

@misc{bitcomp,
  author       = {{NVIDIA}},
  howpublished = {\url{https://developer.nvidia.com/nvcomp}},
  note         = {Online}
}

@inproceedings{zhang2023gpulz,
  title     = {{GPULZ}: Optimizing LZSS Lossless Compression for Multi-byte Data on Modern GPUs},
  author    = {Zhang, Boyuan and Tian, Jiannan and Di, Sheng and Yu, Xiaodong and Swany, Martin and Tao, Dingwen and Cappello, Franck},
  booktitle = {Proceedings of the 37th International Conference on Supercomputing},
  pages     = {348--359},
  year      = {2023}
}

@article{han2022coordnet,
  title     = {Coordnet: Data generation and visualization generation for time-varying volumes via a coordinate-based neural network},
  author    = {Han, Jun and Wang, Chaoli},
  journal   = {IEEE Transactions on Visualization and Computer Graphics},
  year      = {2022},
  publisher = {IEEE}
}

@article{hacc,
  title     = {{HACC}: {E}xtreme scaling and performance across diverse architectures},
  author    = {Habib, Salman and Morozov, Vitali and Frontiere, Nicholas and Finkel, Hal and Pope, Adrian and Heitmann, Katrin and Kumaran, Kalyan and Vishwanath, Venkatram and Peterka, Tom and Insley, Joe and others},
  journal   = {Communications of the ACM},
  volume    = {60},
  number    = {1},
  pages     = {97--104},
  year      = {2016},
  publisher = {ACM}
}

@misc{lcls,
  howpublished = {\url{https://lcls.slac.stanford.edu/lasers/lcls-ii}},
  note         = {Online}
}

@misc{miranda,
  author       = {{Miranda Radiation Hydrodynamics Data}},
  howpublished = {\url{https://wci.llnl.gov/simulation/computer-codes/miranda}},
  note         = {Online},
  year         = {2019}
}

@misc{nyx,
  author       = {{Nyx simulation}},
  howpublished = {\url{https://amrex-astro.github.io/Nyx/}},
  note         = {Online}
}

@misc{qmcpack,
  author       = {{QMCPACK: many-body ab initio Quantum Monte Carlo code}},
  howpublished = {\url{http://vis.computer.org/vis2004contest/data.html}},
  note         = {Online},
  year         = {2019}
}

@article{jhtdb,
  title     = {A public turbulence database cluster and applications to study Lagrangian evolution of velocity increments in turbulence},
  author    = {Li, Yi and Perlman, Eric and Wan, Minping and Yang, Yunke and Meneveau, Charles and Burns, Randal and Chen, Shiyi and Szalay, Alexander and Eyink, Gregory},
  journal   = {Journal of Turbulence},
  number    = {9},
  pages     = {N31},
  year      = {2008},
  publisher = {Taylor \& Francis}
}

@article{rtm,
  author    = {Kayum, Suha and others},
  journal   = {First Break},
  number    = {2},
  pages     = {97--100},
  publisher = {},
  title     = {{GeoDRIVE} -- a high performance computing flexible platform for seismic applications},
  volume    = {38},
  year      = {2020}
}

@misc{miraio,
  author       = {V. Vishwanath, S. Crusan and K. Harms},
  title        = {{Parallel I/O on Mira}},
  howpublished = {\url{https://www.alcf.anl.gov/files/Parallel\_IO\_on\_Mira\_0.pdf}},
  note         = {Online},
  year         = {2019}
}

@inproceedings{tian2021revisiting,
  title     = {{Revisiting Huffman Coding}: Toward Extreme Performance on Modern GPU Architectures},
  author    = {Tian, Jiannan and Rivera, Cody and Di, Sheng and Chen, Jieyang and Liang, Xin and Tao, Dingwen and Cappello, Franck},
  booktitle = {2021 {IEEE} International Parallel and Distributed Processing Symposium
               (IPDPS), Portland, OR, USA, May 17-21, 2021},
  pages     = {881--891},
  publisher = {{IEEE}},
  year      = {2021}
}

@misc{repo-mgardx,
  author       = {},
  title        = {{MGARD-X: A portable implementation of the MGARD lossy compressor supporting various types of GPUs and CPUs}},
  howpublished = {\url{https://github.com/CODARcode/MGARD/blob/master/doc/MGARD-X.md}},
  month        = {},
  year         = {}
}

@misc{repo-sdrbench,
  author       = {{Scientific Data Reduction Benchmarks}},
  howpublished = {\url{https://sdrbench.github.io/}},
  note         = {Online},
  year         = {2019}
}

@misc{testbed-jlse,
  author       = {},
  title        = {Joint Laboratory for System Evaluation – Evaluating future high-performance computing platforms},
  howpublished = {\url{https://www.jlse.anl.gov/}},
  month        = {},
  year         = {},
  note         = {(Accessed on 03/15/2021)}
}

@misc{testbed-thetagpu,
  author       = {},
  title        = {{Theta/ThetaGPU - Argonne Leadership Computing Facility}},
  howpublished = {\url{https://www.alcf.anl.gov/alcf-resources/theta}},
  month        = {},
  year         = {2021},
  note         = {}
}

@article{globus2,
  title     = {Globus platform-as-a-service for collaborative science applications},
  author    = {Ananthakrishnan, Rachana and Chard, Kyle and Foster, Ian and Tuecke, Steven},
  journal   = {Concurrency and Computation: Practice and Experience},
  volume    = {27},
  number    = {2},
  pages     = {290--305},
  year      = {2015},
  publisher = {Wiley Online Library}
}

@inproceedings{cuszplus2021,
  author    = {J. Tian and S. Di and X. Yu and C. Rivera and K. Zhao and S. Jin and Y. Feng and X. Liang and D. Tao and F. Cappello},
  booktitle = {2021 IEEE International Conference on Cluster Computing (CLUSTER)},
  title     = {Optimizing Error-Bounded Lossy Compression for Scientific Data on GPUs},
  year      = {2021},
  pages     = {283-293},
  doi       = {10.1109/Cluster48925.2021.00047},
  publisher = {IEEE Computer Society},
  address   = {Los Alamitos, CA, USA},
  month     = {09}
}

@inproceedings{cusz2020,
  title     = {{{\scshape cuSZ}: An efficient gpu-based error-bounded lossy compression framework for scientific data}},
  author    = {Tian, Jiannan and Di, Sheng and Zhao, Kai and Rivera, Cody and Fulp, Megan Hickman and Underwood, Robert and Jin, Sian and Liang, Xin and Calhoun, Jon and Tao, Dingwen and others},
  booktitle = {Proceedings of the ACM International Conference on Parallel Architectures and Compilation Techniques},
  pages     = {3--15},
  year      = {2020}
}

@inproceedings{liang2018error,
  title     = {Error-controlled lossy compression optimized for high compression ratios of scientific datasets},
  author    = {Liang, Xin and Di, Sheng and Tao, Dingwen and Li, Sihuan and Li, Shaomeng and Guo, Hanqi and Chen, Zizhong and Cappello, Franck},
  booktitle = {2018 IEEE International Conference on Big Data (Big Data)},
  pages     = {438--447},
  year      = {2018},
  publisher = {IEEE},
  address   = {Seattle, WA, USA}
}

@inproceedings{sz17,
  title     = {Significantly improving lossy compression for scientific data sets based on multidimensional prediction and error-controlled quantization},
  author    = {Tao, Dingwen and Di, Sheng and Chen, Zizhong and Cappello, Franck},
  booktitle = {2017 IEEE International Parallel and Distributed Processing Symposium},
  pages     = {1129--1139},
  year      = {2017},
  publisher = {IEEE},
  address   = {Orlando, FL, USA}
}

@inproceedings{zhao2021optimizing,
  title        = {Optimizing Error-Bounded Lossy Compression for Scientific Data by Dynamic Spline Interpolation},
  author       = {Zhao, Kai and Di, Sheng and Dmitriev, Maxim and Tonellot, Thierry-Laurent D. and Chen, Zizhong and Cappello, Franck},
  booktitle    = {2021 IEEE 37th International Conference on Data Engineering (ICDE)},
  pages        = {1643--1654},
  year         = {2021},
  organization = {IEEE}
}

@article{sz3,
  title     = {{SZ3}: A modular framework for composing prediction-based error-bounded lossy compressors},
  author    = {Liang, Xin and Zhao, Kai and Di, Sheng and Li, Sihuan and Underwood, Robert and Gok, Ali M and Tian, Jiannan and Deng, Junjing and Calhoun, Jon C and Tao, Dingwen and others},
  journal   = {IEEE Transactions on Big Data},
  year      = {2022},
  publisher = {IEEE}
}

@inproceedings{qoz,
  title        = {Dynamic quality metric oriented error bounded lossy compression for scientific datasets},
  author       = {Liu, Jinyang and Di, Sheng and Zhao, Kai and Liang, Xin and Chen, Zizhong and Cappello, Franck},
  booktitle    = {2022 SC22: International Conference for High Performance Computing, Networking, Storage and Analysis (SC)},
  pages        = {892--906},
  year         = {2022},
  organization = {IEEE Computer Society}
}

@inproceedings{liu2023faz,
  title     = {{FAZ}: A flexible auto-tuned modular error-bounded compression framework for scientific data},
  author    = {Liu, Jinyang and Di, Sheng and Zhao, Kai and Liang, Xin and Chen, Zizhong and Cappello, Franck},
  booktitle = {Proceedings of the 37th International Conference on Supercomputing},
  pages     = {1--13},
  year      = {2023}
}

@article{FZGPU,
  title   = {{FZ-GPU}: A Fast and High-Ratio Lossy Compressor for Scientific Computing Applications on GPUs},
  author  = {Zhang, Boyuan and Tian, Jiannan and Di, Sheng and Yu, Xiaodong and Feng, Yunhe and Liang, Xin and Tao, Dingwen and Cappello, Franck},
  journal = {arXiv preprint arXiv:2304.12557},
  year    = {2023}
}

@inproceedings{cuszp,
  title     = {{cuSZp}: An Ultra-fast GPU Error-bounded Lossy Compression Framework with Optimized End-to-End Performance},
  author    = {Huang, Yafan and Di, Sheng and Yu, Xiaodong and Li, Guanpeng and Cappello, Franck},
  booktitle = {Proceedings of the International Conference for High Performance Computing, Networking, Storage and Analysis},
  pages     = {1--13},
  year      = {2023}
}

@article{szx,
  title   = {{SZx}: An ultra-fast error-bounded lossy compressor for scientific datasets},
  author  = {Yu, Xiaodong and Di, Sheng and Zhao, Kai and Tao, Dingwen and Liang, Xin and Cappello, Franck and others},
  journal = {arXiv preprint arXiv:2201.13020},
  year    = {2022}
}

@article{hpez,
  title   = {High-performance Effective Scientific Error-bounded Lossy Compression with Auto-tuned Multi-component Interpolation},
  author  = {Liu, Jinyang and Di, Sheng and Zhao, Kai and Liang, Xin and Jin, Sian and Jian, Zizhe and Huang, Jiajun and Wu, Shixun and Chen, Zizhong and Cappello, Franck},
  journal = {arXiv preprint arXiv:2311.12133},
  year    = {2023}
}

@inproceedings{ae-sz,
  title        = {Exploring Autoencoder-based Error-bounded Compression for Scientific Data},
  author       = {Liu, Jinyang and Di, Sheng and Zhao, Kai and Jin, Sian and Tao, Dingwen and Liang, Xin and Chen, Zizhong and Cappello, Franck},
  booktitle    = {2021 IEEE International Conference on Cluster Computing (CLUSTER)},
  pages        = {294--306},
  year         = {2021},
  organization = {IEEE}
}

@inproceedings{SRN-SZ,
  author    = {J. Liu and S. Di and S. Jin and K. Zhao and X. Liang and Z. Chen and F. Cappello},
  booktitle = {2023 IEEE International Conference on Big Data (BigData)},
  title     = {Scientific Error-bounded Lossy Compression with Super-resolution Neural Networks},
  year      = {2023},
  volume    = {},
  issn      = {},
  pages     = {229-236},
  doi       = {10.1109/BigData59044.2023.10386682},
  url       = {https://doi.ieeecomputersociety.org/10.1109/BigData59044.2023.10386682},
  publisher = {IEEE Computer Society},
  address   = {Los Alamitos, CA, USA},
  month     = {dec}
}

@article{underwood2023roibin,
  title     = {{ROIBIN-SZ}: Fast and Science-Preserving Compression for Serial Crystallography},
  author    = {Underwood, Robert and Yoon, Chunhong and Gok, Ali and Di, Sheng and Cappello, Franck},
  journal   = {Synchrotron Radiation News},
  volume    = {36},
  number    = {4},
  pages     = {17--22},
  year      = {2023},
  publisher = {Taylor \& Francis}
}

@misc{globus-compression,
  title         = {Optimizing Scientific Data Transfer on Globus with Error-bounded Lossy Compression},
  author        = {Yuanjian Liu and Sheng Di and Kyle Chard and Ian Foster and Franck Cappello},
  year          = {2023},
  eprint        = {2307.05416},
  archiveprefix = {arXiv},
  primaryclass  = {cs.DC}
}

\end{document}